\documentclass[prd,showpacs,showkeys,nofootinbib,twocolumn]{revtex4-1}
\usepackage{amsmath,amsfonts,amssymb,color}
\usepackage{bm}
\usepackage{pgfplots}
\usepackage[colorlinks=true,urlcolor=blue,linkcolor=blue,citecolor=blue]{hyperref}

\begin{document} 

\title{Generalized Birkhoff theorem in the Poincar\'e gauge gravity theory}

\author{Yuri N. Obukhov}
\email{obukhov@ibrae.ac.ru}
\affiliation{Theoretical Physics Laboratory, Nuclear Safety Institute, 
Russian Academy of Sciences, B.Tulskaya 52, 115191 Moscow, Russia}

\date{ \today}

\begin{abstract}
The analysis of the validity of Birkhoff's theorem about the uniqueness of the spherically symmetric solution of the gravitational field equations is extended to the framework of the Poincar\'e gauge gravity theory. The class of models with the most general Lagrangians of the Yang-Mills type constructed from all possible quadratic invariants of the curvature and the torsion is considered, including both parity-even and parity-odd sectors. We find the families of models in which the weak and strong versions of the generalized Birkhoff theorem are valid, by making use of the double duality technique. 
\end{abstract}


\maketitle

\section{Introduction}\label{intro}

Spherically symmetric solutions are of particular interest in field-theoretic models. In the general relativity (GR) theory, the Schwarzschild metric is a unique solution of Einstein's gravitational field equations under the assumption of a spherical symmetry of spacetime and matter source distribution. This remarkable theoretical result is known as the Birkhoff theorem, although it was first demonstrated independently by Jebsen \cite{Jebsen}, Eiesland \cite{Eiesland}, and Alexandrow \cite{Alexandrow} well before a better known publication of Birkhoff \cite{Birkhoff}. The validity of this theorem is very important, in particular, in view of the fact that the fundamental gravitational experiments in our Solar system are perfectly consistent with the Schwarzschild geometry. For historic overview one may read \cite{Schmidt,Goenner}.

In Einstein's GR with the vanishing cosmological constant, the classic Birkhoff theorem states that the only locally spherically symmetric solution of the vacuum gravitational field equations in four dimensions is the Schwarzschild metric. When the cosmological constant is nontrivial, the Birkhoff theorem is extended to the statement \cite{Schleich,Rindler} that the only locally spherically symmetric solution of Einstein's equations in vacuum is either the Schwarzschild-(anti)de Sitter metric (first obtained by Kottler \cite{Kottler}) or the Nariai \cite{Nariai} (Bertotti-Kasner \cite{Bertotti,Kasner}) spacetime. 

Later the validity of the generalized Birkhoff theorem was established \cite{Riegert,AK,BM,Ellis,Faraoni,Capo,Zegers,DeserF,Wiltshire,Deve,Meng,Dong1,Dong2} for nonvacuum case, for GR in higher dimensions and for a wide variety of alternative gravitational theories (including the conformal, Lovelock, Gauss-Bonnet, Ho\v{r}ava, $F(R)$, and teleparallel models). 

Here we discuss the Poincar\'e gauge (PG) gravity theory which offers a physically meaningful extension of Einstein's GR to the case when the spin of matter is included as a source of the gravitational field along with the mass of matter \cite{Hehl:1976,Ivanenko:1983,Blag:2002,Trautman:2006,yno:2006,yno:2018}. The canonical spin and the energy-momentum currents \cite{WeylBH} underlie the corresponding gauge scheme as the Noether currents for the Poincar\'e group $G = T_4 \rtimes SO(1,3)$ which is the semi-direct product of the 4-parameter group $T_4$ of spacetime translations and the 6-parameter local Lorentz group $SO(1,3)$. In the framework of the consistent Yang-Mills-Kibble-Utiyama field-theoretic approach, the Poincar\'e gauge potentials are identified with the coframe 1-form $\vartheta^\alpha = e^\alpha_i dx^i$ (``translational potential'' corresponding to the $T_4$ subgroup) and the local Lorentz connection 1-form $\Gamma^{\alpha\beta} = -\,\Gamma^{\beta\alpha} = \Gamma_i{}^{\alpha\beta} dx^i$ (``rotational potential'' corresponding to the $SO(1,3)$ subgroup) which give rise to the Riemann-Cartan geometry on the spacetime manifold \cite{MAG,Blag:2013,PBO}. 

We focus here on the analysis of the class of Poincar\'e gauge gravity models based on the general Lagrangians of the Yang--Mills type constructed from all possible quadratic invariants of the curvature and the torsion. In contrast to GR, a spherically symmetric solution is not unique in a general quadratic PG gravity theory. However, certain classes of models do admit the {\it generalized Birkhoff theorem} which can be formulated as follows: the torsion-less  Schwarzschild (Kottler, in general) spacetime is a unique vacuum spherically symmetric solution of the vacuum Poincar\'e gauge field equations. 

This generalized Birkhoff theorem is available in two versions. In the strong version, the spherical symmetry is understood as the form-invariance of physical and geometrical variables under the $SO(3)$ group of rotations, whereas in the weak $O(3)$ version one assumes the invariance under the rotations {\it and} spatial reflections. The earlier work was confined to the parity-even Lagrangians only \cite{Rama,RauchGRG:1982,Rauch:1981,RauchPRD:1982,Neville:1978,Neville:1980,Cruz,OPZ}. 

Following the lines of \cite{OPZ}, we now extend the consideration and generalize the class of quadratic PG models to include the parity-odd sector. In the recent times, there is a growing interest to such interactions \cite{OPZ,HoN,Diakonov,Baekler1,Baekler2,Kara,BC}. Quite generally, there are no convincing experimental evidence or compelling theoretical arguments which could rule out the violation of parity in gravity. On the contrary, the discovery of $P$ and $CP$ nonconservation in the weak interaction processes \cite{CP} had stimulated considerable efforts in the experimental search for the electric dipole moments of elementary particles \cite{POS}, and the possibility of extending the gravitational Lagrangian by parity odd terms has been proposed already in the mid of 1960's \cite{LO}. Later such extensions were widely studied in the context of the classical and quantum gravity theory \cite{Hari,Hoj,MH}, in particular in Ashtekar's approach and loop quantum gravity \cite{Holst,CK}. Moreover, the inclusion of parity-nonconserving terms appears to be important for the discussion of such fundamental physical issues as the baryon asymmetry of the universe (predicted by Sakharov \cite{ADS}), where the parity odd terms can be induced by the quantum vacuum structure \cite{Randono,Pop,Bj}. 

In order to make the discussion as transparent as possible, all the physical and geometrical objects related to the parity-odd sector (such as coupling constants, irreducible pieces of the curvature and the torsion, etc.) are marked by an overline, to distinguish them from the corresponding parity-even objects. Other basic notation and conventions are consistent with \cite{MAG,Blag:2013}. In particular, Greek indices $\alpha, \beta, \dots = 0, \dots, 3$, denote the anholonomic components (for example, of a coframe $\vartheta^\alpha$), while the Latin indices $i,j,\dots =0,\dots, 3$, label the holonomic components (e.g., $dx^i$). The anholonomic vector frame basis $e_\alpha$ is dual to the coframe basis in the sense that $e_\alpha\rfloor\vartheta^\beta = \delta_\alpha^\beta$, where $\rfloor$ denotes the interior product. The volume 4-form is denoted by $\eta$, and the $\eta$-basis in the space of exterior forms is constructed with the help of the interior products as $\eta_{\alpha_1 \dots\alpha_p}:= e_{\alpha_p}\rfloor\dots e_{\alpha_1}\rfloor\eta$, $p=1,\dots,4$. They are related to the $\vartheta$-basis via the Hodge dual operator $^*$, for example, $\eta_{\alpha\beta} = {}^*\!\left(\vartheta_\alpha\wedge\vartheta_\beta\right)$. The Minkowski metric is $g_{\alpha\beta} = {\rm diag}(c^2,-1,-1,-1)$. For exterior forms $\omega$ of various type (which in general may carry extra anholonomic indices) we use the standard notation for the Lie derivative $L_\xi\omega = d(\xi\rfloor\omega) + \xi\rfloor(d\omega)$ along a vector field $\xi$.

\section{Poincar\'e gauge field equations}

The torsion 2-form $T^\alpha = D\vartheta^\alpha = d\vartheta^\alpha +\Gamma_\beta{}^\alpha\wedge\vartheta^\beta$ can be decomposed into the 3 irreducible parts, whereas the curvature 2-form $R^{\alpha\beta} = d\Gamma^{\alpha\beta} + \Gamma_\gamma{}^\beta\wedge\Gamma^{\alpha\gamma}$ has 6 irreducible pieces. Their definition is presented in Appendix~\ref{appA}. These irreducible parts of the Poincar\'e gauge field strengths are the building blocks for constructing the Yang-Mills type Lagrangian of the gravitational field.

The general quadratic model is described by the Lagrangian 4-form that contains all possible linear and quadratic invariants of the torsion and the curvature:
\begin{eqnarray}
V &=& {\frac {1}{2\kappa c}}\Big\{\Big(a_0\eta_{\alpha\beta} + \overline{a}_0
\vartheta_\alpha\wedge\vartheta_\beta\Big)\wedge R^{\alpha\beta} - 2\lambda_0\eta \nonumber\\
&& -\,T^\alpha\wedge\sum_{I=1}^3\left[a_I\,{}^*({}^{(I)}T_\alpha) + \overline{a}_I
\,{}^{(I)}T_\alpha\right]\Big\} \nonumber\\ &&
- \,{\frac 1{2\rho}}R^{\alpha\beta}\wedge\sum_{I=1}^6 \left[b_I\,{}^*({}^{(I)}\!R_{\alpha\beta}) 
+ \overline{b}_I\,{}^{(I)}\!R_{\alpha\beta}\right].\label{LRT}
\end{eqnarray}
The structure of quadratic part of the Lagrangian is specified by the set of coupling constants: $\rho$, $a_1, a_2, a_3$, $b_1, \cdots, b_6$ and $\overline{a}_1, \overline{a}_2, \overline{a}_3$, $\overline{b}_1, \cdots, \overline{b}_6$. In the parity-odd sector, not all constants are independent because some of terms in (\ref{LRT}) are the same \cite{yno:2018}, and accordingly we have to put $\overline{a}_2 = \overline{a}_3$, $\overline{b}_2 = \overline{b}_4$ and $\overline{b}_3 = \overline{b}_6$. The coupling constants $a_I$, $\overline{a}_I$, $b_I$ and $\overline{b}_I$ are dimensionless, whereas the dimension $[{\frac 1\rho}] = [\hbar]$. By demanding the existence of a macroscopic limit to GR, we identify $\kappa = 8\pi G/c^4$ as Einstein's gravitational constant. 

The Lagrangian (\ref{LRT}) has a clear structure: the first line encompasses the terms {\em linear} in the curvature, the second line contains the {\em torsion quadratic} terms (all of which have the same dimension of an area $[\ell^2]$), and finally the third line contains the {\it curvature quadratic} invariants. For completeness, the cosmological constant is included (with the dimension of an inverse area, $[\lambda_0] = [\ell^{-2}]$). A special case $a_0 = 0$ and $\overline{a}_0 = 0$ describes the purely quadratic model without the Hilbert-Einstein linear term in the Lagrangian. Such models describe the microscopic gravitational phenomena which are naturally characterized by the parameter 
\begin{equation}
\ell_\rho^2 = {\frac {\kappa c}{\rho}}\label{lr}
\end{equation}
with the dimension of an area $[\ell_\rho^2] = [\ell^2]$. In order to provide a consistent macroscopic limit to Einstein's GR (that is solidly confirmed at the large distances), we assume that $a_0 \neq 0$. 

The {\it vacuum} Poincar\'e gravity field equations are obtained from the variation of the gravitational action $\int V$ with respect to the translational and rotational gauge potentials: 
\begin{eqnarray}
{\mathcal E}_\alpha = {\frac {\delta V}{\delta\vartheta^\alpha}} = {\frac {a_0}2}\eta_{\alpha\beta\gamma}
\wedge R^{\beta\gamma} + \overline{a}_0R_{\alpha\beta}\wedge\vartheta^\beta - \lambda_0\eta_\alpha  \nonumber\\
+ \, q^{(T)}_\alpha + \ell_\rho^2\,q^{(R)}_\alpha - Dh_\alpha = 0,\label{ERT1}\\
{\mathcal C}_{\alpha\beta} = {\frac {\delta V}{\delta\Gamma^{\alpha\beta}}} = a_0\,\eta_{\alpha\beta\gamma}
\wedge T^{\gamma} + 2\overline{a}_0\,T_{[\alpha}\wedge\vartheta_{\beta]}  \nonumber\\
+ \,2h_{[\alpha}\wedge\vartheta_{\beta]} - 2\ell_\rho^2\,Dh_{\alpha\beta} = 0.\label{ERT2}
\end{eqnarray}
Here we introduced the 2-forms which are linear functions of the torsion and the curvature, respectively, by
\begin{eqnarray}
h_\alpha &=& \sum_{I=1}^3\left[a_I\,{}^*({}^{(I)}T_\alpha) 
+ \overline{a}_I\,{}^{(I)}T_\alpha\right],\label{hT}\\
h_{\alpha\beta} &=& \sum_{I=1}^6\left[b_I\,{}^*({}^{(I)}\!R_{\alpha\beta}) 
+ \overline{b}_I\,{}^{(I)}\!R_{\alpha\beta}\right],\label{hR}
\end{eqnarray}
and the 3-forms which are quadratic in the torsion and in the curvature, respectively:
\begin{eqnarray}
q^{(T)}_\alpha &=& {\frac 12}\left[(e_\alpha\rfloor T^\beta)\wedge h_\beta - T^\beta\wedge 
e_\alpha\rfloor h_\beta\right],\label{qaT}\\
q^{(R)}_\alpha &=& {\frac 12}\left[(e_\alpha\rfloor R^{\beta\gamma})\wedge h_{\beta\gamma} 
- R^{\beta\gamma}\wedge e_\alpha\rfloor h_{\beta\gamma}\right].\label{qaR}
\end{eqnarray}

\section{Spherically symmetric fields in Poincar\'e gauge gravity}

The analysis of the validity of the generalized Birkhoff theorem in PG is based on the appropriate ansatz for the coframe $\vartheta^\alpha$ and the local Lorentz connection $\Gamma^{\alpha\beta}$. Let us choose the local coordinates $x^i = (t, r, \theta, \varphi)$. The most general spherically symmetric spacetime interval reads
\begin{equation}
ds^2 = A^2dt^2 - B^2dr^2 - C^2(d\theta^2 + \sin^2\theta d\varphi^2),\label{Sds}
\end{equation}
where $A = A(t,r)$, $B = B(t,r)$, $C = C(t,r)$ are arbitrary functions of time and radial coordinate.

This follows from the study of the invariance under the action of the rotation symmetry group on the spacetime manifold. The corresponding infinitesimal motion is generated by the vector fields
\begin{eqnarray}
\xi_x &=& \sin\varphi\,\partial_\theta + {\frac {\cos\varphi}{\sin\theta}}\cos\theta\,\partial_\varphi,\label{xix} \\
\xi_y &=& -\,\cos\varphi\,\partial_\theta + {\frac {\sin\varphi}{\sin\theta}}\cos\theta\,\partial_\varphi,\label{xiy} \\
\xi_z &=& -\,\partial_\varphi,\label{xiz}
\end{eqnarray}
the commutators of which form the Lie algebra $so(3)$ of the rotation group. The form (\ref{Sds}) of the line element $ds^2 = g_{ij}dx^idx^j$ is then fixed by demanding the invariance of the metric under the action $\xi: SO(3)\times M \rightarrow M$ of the group $SO(3)$ on the spacetime manifold $M$. Technically, the invariance condition means the vanishing of the Lie derivative $L_\xi\,g_{ij} = 0$ along the vector fields (\ref{xix})-(\ref{xiz}). For more details see, e.g., \cite{Mink,Sur,MH1,MH2}.

The general spherically symmetric configuration of the Poincar\'e gauge fields is described by the coframe $\vartheta^\alpha$:
\begin{equation}\label{cofSS}
\vartheta^{\hat 0} = Adt,\ \vartheta^{\hat 1} = Bdr,\
\vartheta^{\hat 2} = Cd\theta,\ \vartheta^{\hat 3} = C\sin\theta d\varphi,
\end{equation}
and the local Lorentz connection $\Gamma^{\alpha\beta}$:
\begin{align}\label{G01S}
\Gamma^{{\hat 0}{\hat 1}} =& fdt + gdr, &
\Gamma^{{\hat 2}{\hat 3}} =& \overline{f}dt + \overline{g}dr - \cos\theta d\varphi,\\
\Gamma^{{\hat 0}{\hat 2}} =& pd\theta - \overline{q}\sin\theta d\varphi, &
\Gamma^{{\hat 3}{\hat 1}} =& \overline{p}d\theta - q\sin\theta d\varphi,\label{G02S}\\
\Gamma^{{\hat 0}{\hat 3}} =& p\sin\theta d\varphi + \overline{q}d\theta, &
\Gamma^{{\hat 1}{\hat 2}} =& \overline{p}\sin\theta d\varphi + qd\theta.\label{G03S}
\end{align}
The configuration for the local Lorentz connection encompasses the eight functions $f = f(t,r)$, $g = g(t,r)$, $p = p(t,r)$, $q = q(t,r)$, and $\overline{f} = \overline{f}(t,r)$, $\overline{g} = \overline{g}(t,r)$, $\overline{p} = \overline{p}(t,r)$, $\overline{q} = \overline{q}(t,r)$.

The Poincar\'e gauge potentials (\ref{cofSS}) and (\ref{G01S})-(\ref{G03S}) satisfy the invariance conditions
\begin{eqnarray}
L_\xi\,\vartheta^\alpha = {\stackrel {(\xi)}{\lambda}}{}_\beta{}^\alpha\,\vartheta^\beta,\qquad
L_\xi\,\Gamma^{\alpha\beta} = -\,D{\stackrel {(\xi)}{\lambda}}{}^{\alpha\beta}.\label{LieD}
\end{eqnarray}
Here the Lie algebra-valued ${\stackrel {(\xi)}{\lambda}}{}^{\alpha\beta} = -\,{\stackrel {(\xi)}{\lambda}}{}^{\beta\alpha}$ parameter is determined by vector fields which generate symmetries. Explicitly, for the rotation symmetry (\ref{xix})-(\ref{xiz}) we have:
\begin{eqnarray}
{\stackrel {(\xi_x)}{\lambda}}{}^{\alpha\beta} &=& 2{\frac {\cos\varphi}{\sin\theta}}
\,\delta^{[\alpha}_{\hat 2}\delta^{\beta]}_{\hat 3},\label{lx}\\
{\stackrel {(\xi_y)}{\lambda}}{}^{\alpha\beta} &=& 2{\frac {\sin\varphi}{\sin\theta}}
\,\delta^{[\alpha}_{\hat 2}\delta^{\beta]}_{\hat 3},\label{ly}\\
{\stackrel {(\xi_z)}{\lambda}}{}^{\alpha\beta} &=& 0.\label{lz}
\end{eqnarray}

The Poincar\'e gauge fields (\ref{cofSS}) and (\ref{G01S})-(\ref{G03S}) represent a general solution of the invariance conditions (\ref{LieD}). Obviously, these configurations are determined up to an arbitrary local Lorentz transformation $\Lambda_\alpha{}^\beta = \Lambda_\alpha{}^\beta(x) \in SO(1,3)$ of the coframe, dual frame and connection
\begin{eqnarray}\label{cofL}
\vartheta^\mu &\rightarrow& \Lambda_\alpha{}^\mu\vartheta^\alpha,\qquad
e_\alpha \rightarrow \Lambda^{-1}{}_\alpha{}^\mu e_\mu,\\
\Gamma_\alpha{}^\beta &\rightarrow& \Lambda_\nu{}^\beta\Gamma_\mu{}^\nu\Lambda^{-1}{}_\alpha{}^\mu + 
\Lambda_\mu{}^\beta d\Lambda^{-1}{}_\alpha{}^\mu,\label{connL}
\end{eqnarray}
however, without loosing generality it is more convenient to work with the diagonal coframe (\ref{cofSS}). For the transformed configurations (\ref{cofL}), (\ref{connL}), the Lie algebra valued parameter ${\stackrel {(\xi)}{\lambda}}{}^{\alpha\beta}$ is accordingly transformed.

It is important to notice that the general spherically symmetric configuration (\ref{cofSS}) and (\ref{G01S})-(\ref{G03S}) is only invariant under the group of proper rotations $SO(3)$, whereas this configuration in not invariant under spatial reflections. This is manifest by the presence of parity-odd terms with nontrivial functions $\overline{f}$, $\overline{g}$, $\overline{p}$, $\overline{q}$. For the thorough discussion of reflections, see \cite{MH2}. When in addition to the pure rotations we demand also the invariance under reflections, the symmetry group is extended to the full rotation group $O(3)$. Such an extension imposes an additional condition on field configurations, which forbids parity-odd terms: $\overline{f} = \overline{g} = \overline{p} = \overline{q} = 0$. As a result, the number of arbitrary functions in (\ref{G01S})-(\ref{G03S}) is reduced from eight to four. 

There exists an alternative approach, in which the symmetries are discussed directly for the tensor field configurations such as the torsion and the curvature. Our study is completely consistent with the latter, since in view of the invariance condition (\ref{LieD}) we straightforwardly find, with $\xi = \xi^i\partial_i$,
\begin{eqnarray}
L_\xi g_{ij} &=& (\partial_i\xi^k)\,g_{kj} + (\partial_j\xi^k)\,g_{ik} = 0,\label{dg}\\ 
L_\xi\Gamma_{ki}{}^j &=&  (\partial_k\xi^l)\,\Gamma_{li}{}^j + (\partial_i\xi^l)\,\Gamma_{kl}{}^j - (\partial_l\xi^j)\,\Gamma_{ki}{}^l\nonumber\\
&& + \,\partial^2_{ki}\xi^j = 0,\label{dG}
\end{eqnarray}
for the metric $g_{ij} = e_i{}^\alpha e_j{}^\beta g_{\alpha\beta}$ and the world connection $\Gamma_{ki}{}^j = e^\alpha_i\,\Gamma_{k\alpha}{}^\beta\,e_\beta^j + e^\alpha_i\partial_k\,e_\alpha^j$. As a result, all the relevant tensor fields constructed from the metric and connection are automatically invariant: $L_\xi T_{ki}{}^j = 0$, $L_\xi R_{kli}{}^j = 0$, and so on.

Here we use a more fundamental approach, starting from the gravitational field potentials $(\vartheta^\alpha, \Gamma^{\alpha\beta})$.

\section{Spherically symmetric curvature and torsion}

From the gauge gravitational potentials (\ref{cofSS}) and (\ref{G01S})-(\ref{G03S}) we can straightforwardly compute the Poincar\'e gauge field strengths: the curvature (``rotational'' field strength) and the torsion (``translational'' field strength). As a preliminary technical step, it is convenient to divide the anholonomic indices, $\alpha,\beta,\dots$, into the two groups: $A,B,\dots = 0,1$ and $a,b,\dots = 2,3$.

Then for the components of the Riemann-Cartan curvature 2-form we find
\begin{eqnarray}
R^{AB} &=& \mu\,\vartheta^A\wedge\vartheta^B + \overline{\mu}\,{}^*\!(\vartheta^A\wedge\vartheta^B),\label{RABS}\\
R^{ab} &=& \nu\,\vartheta^a\wedge\vartheta^b + \overline{\nu}\,{}^*\!(\vartheta^a\wedge\vartheta^b),\label{RabS}\\
R^{Ab} &=& -\,R^{bA} = U^A\wedge\vartheta^b + {}^*\!(\overline{U}^A\wedge\vartheta^b),\label{RAbS}
\end{eqnarray}
where we denoted
\begin{align}
\mu &= {\frac {-f' + \dot{g}}{AB}},& 
\overline{\mu} &= 2{\frac {p\overline{p} + q\overline{q}}{C^2}},\label{mumuS}\\
\nu &= {\frac {p^2 - \overline{p}^2 - q^2 + \overline{q}^2 + 1}{C^2}},&
\overline{\nu} &= {\frac {- \overline{f}\,' + \dot{\overline{g}}}{AB}},\label{nunuS}
\end{align}
and introduced the 1-forms $U^A$ and $\overline{U}{}^A$ with the components
\begin{eqnarray}
U^{\hat 0} &=& {\frac {\dot{p} + fq + \overline{f}\,\overline{q}}{AC}}\,\vartheta^{\hat 0}
+ {\frac {p' + gq + \overline{g}\,\overline{q}}{BC}}\,\vartheta^{\hat 1}\,,\label{U0S}\\
U^{\hat 1} &=& {\frac {\dot{q} + fp - \overline{f}\,\overline{p}}{AC}}\,\vartheta^{\hat 0}
+ {\frac {q' + gp - \overline{g}\,\overline{p}}{BC}}\,\vartheta^{\hat 1}\,,\label{U1S}\\
\overline{U}{}^{\hat 0} &=&  {\frac {-\,\overline{q}\,' + g\overline{p} + \overline{g}p}{BC}}
\,\vartheta^{\hat 0} + {\frac {-\,\dot{\overline{q}} + f\overline{p} + \overline{f}p}{AC}}
\,\vartheta^{\hat 1},\label{Ut0S}\\
\overline{U}{}^{\hat 1} &=&  {\frac {\overline{p}\,' - g\overline{q} + \overline{g}q}{BC}}
\,\vartheta^{\hat 0} + {\frac {\dot{\overline{p}} - f\overline{q} + \overline{f}q}{AC}}
\,\vartheta^{\hat 1}.\label{Ut1S}
\end{eqnarray}
Hereafter, the dot $\dot{}$ and the prime $^\prime$ denote derivatives with respect to the time $t$ and the radial coordinate $r$, respectively.

Using these objects, we can construct the Yang-Mills type quadratic curvature invariant
\begin{eqnarray}
R^{\alpha\beta}\wedge{}^*R_{\alpha\beta} &=& 2\,(\nu^2 - \mu^2 + \overline{\mu}^2 - \overline{\nu}^2)\,\eta 
\nonumber\\
&& + \,4(U^A\wedge{}^*{}U_A - \overline{U}{}^A\wedge{}^*{}\overline{U}_A).\label{RRS}
\end{eqnarray}
Raising and lowering of the indices is done with the help of the corresponding effective two-dimensional metric tensors which arise as the sub-blocks of the four-dimensional Minkowski metric $g_{\alpha\beta} = \left(\begin{array}{cc}g_{AB} & 0 \\ 0 & g_{ab}\end{array}\right)$:
\begin{equation}
g_{AB} = \left(\begin{array}{cc}1 & 0 \\ 0 & -1\end{array}\right),\quad
g_{ab} = \left(\begin{array}{cc}-1 & 0 \\ 0 & -1\end{array}\right).\label{met2}
\end{equation}

Similarly, a direct computation yields for components of the spherically symmetric torsion 2-form:
\begin{eqnarray}
T^A = \vartheta^A\wedge (2V + T/3) - {}^\ast(\vartheta^A\wedge (2\overline{V} + \overline{T}/3)),\label{T1}\\
T^a = -\,\vartheta^a\wedge (V - T/3) + {}^\ast(\vartheta^a\wedge (\overline{V} - \overline{T}/3)).\label{T2}
\end{eqnarray}
Here we introduced the quartet of 1-forms
\begin{equation}
T = t_A\vartheta^A,\  \overline{T} = \overline{t}_A\vartheta^A,\quad 
V = v_A \vartheta^A,\  \overline{V} = \overline{v}_A\vartheta^A,\label{VVTT}
\end{equation}
the components of which read explicitly as follows:
\begin{eqnarray}
t_{\hat 0} &=& {\frac 1A}\left(-\,{\frac {\dot{B}}{B}} - 2{\frac {\dot{C}}{C}}
+ g\,{\frac {A}{B}} + 2p\,{\frac {A}{C}}\right),\label{u0S}\\
t_{\hat 1} &=& {\frac 1B}\left(-\,{\frac {A'}{A}} - 2{\frac {C'}{C}}
+ f\,{\frac {B}{A}} - 2q\,{\frac {B}{C}}\right),\label{u1S}\\
\overline{t}{}_{\hat 0} &=& 2{\frac {\overline{g}}{B}} + 4{\frac {\overline{p}}{C}},\qquad
\overline{t}{}_{\hat 1} = 2{\frac {\overline{f}}{A}} + 4{\frac {\overline{q}}{C}},\label{tuS}
\end{eqnarray}
and 
\begin{eqnarray}
v_{\hat 0} &=& {\frac 1{3A}}\left(-\,{\frac {\dot{B}}{B}} + {\frac {\dot{C}}{C}}
+ g\,{\frac {A}{B}} - p\,{\frac {A}{C}}\right),\label{v0S}\\
v_{\hat 1} &=& {\frac 1{3B}}\left(-\,{\frac {A'}{A}} + {\frac {C'}{C}}
+ f\,{\frac {B}{A}} + q\,{\frac {B}{C}}\right),\label{v1S}\\
\overline{v}{}_{\hat 0} &=& -\,{\frac {\overline{g}}{3B}} + {\frac {\overline{p}}{3C}},\qquad
\overline{v}{}_{\hat 1} = -\,{\frac {\overline{f}}{3A}} + {\frac {\overline{q}}{3C}}.\label{tvS}
\end{eqnarray}
All together, the general spherically symmetric torsion configuration thus includes eight variables (\ref{u0S})-(\ref{tvS}) -- the components of the 1-forms $T, \overline{T}, V, \overline{V}$:
\begin{equation}
t_A(t,r),\ \overline{t}_A(t,r),\ v_A(t,r),\ \overline{v}_A(t,r),\qquad A = 0,1.\label{uv}
\end{equation}

These four 1-forms determine the torsion invariants, in particular for the Yang-Mills quadratic invariant we find
\begin{eqnarray}
T^\alpha\wedge{}^*T_\alpha &=& 6\left(V\wedge{}^*{}V - \overline{V}\wedge{}^*{}\overline{V}\right)\nonumber\\
&& +\,{\frac 13}\left(T\wedge{}^*{}T - \overline{T}\wedge{}^*{}\overline{T}\right).\label{TTS}
\end{eqnarray}

For the analysis of the validity of the generalized Birkhoff theorem, it is instructive to find the structure of irreducible parts of the spherically symmetric torsion and curvature 2-forms.

\subsection{Irreducible torsion decomposition}

The decomposition of the spherically symmetric torsion (\ref{T1})-(\ref{T2}) into its three irreducible parts is straightforward. The 1st irreducible (traceless tensor) torsion is constructed from the 1-forms $V$ and $\overline{V}$:
\begin{eqnarray}
{}^{(1)}\!T^A &=& 2\vartheta^A\wedge V + 2e^A\rfloor{}^\ast\overline{V},\label{ST1}\\
{}^{(1)}\!T^a &=& -\,\vartheta^a\wedge V - e^a\rfloor{}^\ast\overline{V},\label{ST2}
\end{eqnarray}
whereas the 2nd (torsion trace) and the 3rd (axial trace) parts read
\begin{eqnarray}
{}^{(2)}\!T^\alpha = {\frac 13}\vartheta^\alpha\wedge T,&\qquad&
{}^{(3)}\!T^\alpha = {\frac 13}e^\alpha\rfloor {}^\ast\overline{T}.\label{ST23}
\end{eqnarray}

\subsection{Irreducible curvature decomposition}

All six irreducible parts of the curvature are nontrivial, in general. In order to clarify the structure of the irreducible pieces of the spherically symmetric curvature 2-form (\ref{RABS})-(\ref{RAbS}), one needs a somewhat different parametrization of its components. Namely, it turns out to be more convenient to replace $\mu, \nu, \overline{\mu}, \overline{\nu}$ and components of the 1-forms $U^A$ and $\overline{U}{}^A$ by the set of the new variables as follows:
\begin{widetext}
\begin{eqnarray}
\mu = {\frac {{\mathcal M} + 2{\mathcal K} + {\mathcal N}}{2}},\qquad
\nu = {\frac {{\mathcal M} + 2{\mathcal K} - {\mathcal N}}{2}}, &\qquad&
\overline{\mu} = {\frac {\overline{\mathcal M} + 2\overline{\mathcal K} + \overline{\mathcal N}}{2}},\qquad
\overline{\nu} = {\frac {\overline{\mathcal M} + 2\overline{\mathcal K} - \overline{\mathcal N}}{2}},\label{MN}\\
U^{\hat{0}}_{\hat{0}} = {\frac {{\mathcal M} - {\mathcal K} + {\mathcal L}}{2}},\qquad
U^{\hat{0}}_{\hat{1}} = {\frac {{\mathcal Q} + {\mathcal P}}{2}},&\qquad& 
U^{\hat{1}}_{\hat{0}} = {\frac {{\mathcal Q} - {\mathcal P}}{2}},\qquad
U^{\hat{1}}_{\hat{1}} = {\frac {{\mathcal M} - {\mathcal K} - {\mathcal L}}{2}}.\label{LP}\\
\overline{U}{}^{\hat{0}}_{\hat{0}} = {\frac {\overline{\mathcal M} - \overline{\mathcal K} + \overline{\mathcal L}}{2}},\qquad
\overline{U}{}^{\hat{0}}_{\hat{1}} = {\frac {\overline{\mathcal Q} + \overline{\mathcal P}}{2}},&\qquad& 
\overline{U}{}^{\hat{1}}_{\hat{0}} = {\frac {\overline{\mathcal Q} - \overline{\mathcal P}}{2}},\qquad
\overline{U}{}^{\hat{1}}_{\hat{1}} = {\frac {\overline{\mathcal M} - \overline{\mathcal K} - \overline{\mathcal L}}{2}}.\label{OLP}
\end{eqnarray}
In terms of the new variables, one then finds the 1st irreducible (the Weyl) part of the curvature:
\begin{eqnarray}
{}^{(1)}\!R^{\hat{0}\hat{1}} = {\mathcal K}\,\vartheta^{\hat{0}}\wedge\vartheta^{\hat{1}}
- \overline{\mathcal K}\,\vartheta^{\hat{2}}\wedge\vartheta^{\hat{3}},\quad
{}^{(1)}\!R^{\hat{0}\hat{2}} = -\,{\frac {{\mathcal K}}{2}}\,\vartheta^{\hat{0}}\wedge\vartheta^{\hat{2}}
+ {\frac {\overline{\mathcal K}}{2}}\,\vartheta^{\hat{3}}\wedge\vartheta^{\hat{1}},\quad
{}^{(1)}\!R^{\hat{0}\hat{3}} = -\,{\frac {{\mathcal K}}{2}}\,\vartheta^{\hat{0}}\wedge\vartheta^{\hat{3}}
+ {\frac {\overline{\mathcal K}}{2}}\,\vartheta^{\hat{1}}\wedge\vartheta^{\hat{2}},\label{1R0a}\\
{}^{(1)}\!R^{\hat{2}\hat{3}} = {\mathcal K}\,\vartheta^{\hat{2}}\wedge\vartheta^{\hat{3}}
+ \overline{\mathcal K}\,\vartheta^{\hat{0}}\wedge\vartheta^{\hat{1}},\quad
{}^{(1)}\!R^{\hat{3}\hat{1}} = -\,{\frac {{\mathcal K}}{2}}\,\vartheta^{\hat{3}}\wedge\vartheta^{\hat{1}}
- {\frac {\overline{\mathcal K}}{2}}\,\vartheta^{\hat{0}}\wedge\vartheta^{\hat{2}},\quad
{}^{(1)}\!R^{\hat{1}\hat{2}} = -\,{\frac {{\mathcal K}}{2}}\,\vartheta^{\hat{1}}\wedge\vartheta^{\hat{2}}
- {\frac {\overline{\mathcal K}}{2}}\,\vartheta^{\hat{0}}\wedge\vartheta^{\hat{3}},\label{1Rab}
\end{eqnarray}
the 2nd (the pair anticommutator) part:
\begin{eqnarray}
{}^{(2)}\!R^{\hat{0}\hat{1}} = -\,{\frac {\overline{\mathcal N}}{2}}\,\vartheta^{\hat{2}}\wedge\vartheta^{\hat{3}},\qquad
{}^{(2)}\!R^{\hat{0}\hat{2}} = -\,{\frac {\overline{\mathcal L}}{2}}\,\vartheta^{\hat{3}}\wedge\vartheta^{\hat{1}}
+ {\frac {\overline{\mathcal P}}{2}}\,\vartheta^{\hat{0}}\wedge\vartheta^{\hat{3}},\qquad
{}^{(2)}\!R^{\hat{0}\hat{3}} = -\,{\frac {\overline{\mathcal L}}{2}}\,\vartheta^{\hat{1}}\wedge
\vartheta^{\hat{2}} - {\frac {\overline{\mathcal P}}{2}}\,\vartheta^{\hat{0}}\wedge
\vartheta^{\hat{2}},\label{2R0a}\\
{}^{(2)}\!R^{\hat{2}\hat{3}} = -\,{\frac {\overline{\mathcal N}}{2}}\,\vartheta^{\hat{0}}\wedge\vartheta^{\hat{1}},\qquad
{}^{(2)}\!R^{\hat{3}\hat{1}} = -\,{\frac {\overline{\mathcal L}}{2}}\,\vartheta^{\hat{0}}\wedge\vartheta^{\hat{2}}
- {\frac {\overline{\mathcal P}}{2}}\,\vartheta^{\hat{1}}\wedge\vartheta^{\hat{2}},\qquad 
{}^{(2)}\!R^{\hat{1}\hat{2}} = -\,{\frac {\overline{\mathcal L}}{2}}\,\vartheta^{\hat{0}}\wedge\vartheta^{\hat{3}}
+ {\frac {\overline{\mathcal P}}{2}}\,\vartheta^{\hat{3}}\wedge\vartheta^{\hat{1}},\label{2Rab}
\end{eqnarray}
the 4th (the traceless Ricci) part: 
\begin{eqnarray}
{}^{(4)}\!R^{\hat{0}\hat{1}} =  {\frac {{\mathcal N}}{2}}\,\vartheta^{\hat{0}}\wedge\vartheta^{\hat{1}},\qquad
{}^{(4)}\!R^{\hat{0}\hat{2}} = {\frac {{\mathcal L}}{2}}\,\vartheta^{\hat{0}}\wedge\vartheta^{\hat{2}}
+ {\frac {{\mathcal P}}{2}}\,\vartheta^{\hat{1}}\wedge\vartheta^{\hat{2}},\qquad
{}^{(4)}\!R^{\hat{0}\hat{3}} =  {\frac {{\mathcal L}}{2}}\,\vartheta^{\hat{0}}\wedge
\vartheta^{\hat{3}} - {\frac {{\mathcal P}}{2}}\,\vartheta^{\hat{3}}\wedge
\vartheta^{\hat{1}},\label{4R0a}\\
{}^{(4)}\!R^{\hat{2}\hat{3}} = -\,{\frac {{\mathcal N}}{2}}\,\vartheta^{\hat{2}}\wedge\vartheta^{\hat{3}},\qquad
{}^{(4)}\!R^{\hat{3}\hat{1}} = -\,{\frac {{\mathcal L}}{2}}\,\vartheta^{\hat{3}}\wedge\vartheta^{\hat{1}}
+ {\frac {{\mathcal P}}{2}}\,\vartheta^{\hat{0}}\wedge\vartheta^{\hat{3}},\qquad 
{}^{(4)}\!R^{\hat{1}\hat{2}} = -\,{\frac {{\mathcal L}}{2}}\,\vartheta^{\hat{1}}\wedge\vartheta^{\hat{2}}
- {\frac {{\mathcal P}}{2}}\,\vartheta^{\hat{0}}\wedge\vartheta^{\hat{2}},\label{4Rab}
\end{eqnarray}
the 5th (the skew-symmetric Ricci) part:
\begin{eqnarray}\label{5R0a}
{}^{(5)}\!R^{\hat{0}\hat{1}} = 0,\qquad
{}^{(5)}\!R^{\hat{0}\hat{2}} = {\frac {{\mathcal Q}}{2}}\,\vartheta^{\hat{1}}\wedge\vartheta^{\hat{2}}
+ {\frac {\overline{\mathcal Q}}{2}}\,\vartheta^{\hat{0}}\wedge\vartheta^{\hat{3}},\qquad
{}^{(5)}\!R^{\hat{0}\hat{3}} = -\,{\frac {{\mathcal Q}}{2}}\,\vartheta^{\hat{3}}\wedge\vartheta^{\hat{1}}
- {\frac {\overline{\mathcal Q}}{2}}\,\vartheta^{\hat{0}}\wedge\vartheta^{\hat{2}}, \\
{}^{(5)}\!R^{\hat{2}\hat{3}} = 0,\qquad 
{}^{(5)}\!R^{\hat{3}\hat{1}} = -\,{\frac {{\mathcal Q}}{2}}\,\vartheta^{\hat{0}}\wedge\vartheta^{\hat{3}}
+ {\frac {\overline{\mathcal Q}}{2}}\,\vartheta^{\hat{1}}\wedge\vartheta^{\hat{2}},\qquad 
{}^{(5)}\!R^{\hat{1}\hat{2}} = {\frac {{\mathcal Q}}{2}}\,\vartheta^{\hat{0}}\wedge\vartheta^{\hat{2}}
- {\frac {\overline{\mathcal Q}}{2}}\,\vartheta^{\hat{3}}\wedge\vartheta^{\hat{1}},\label{5Rab}
\end{eqnarray}
and finally the 3rd and 6th (the curvature pseudoscalar and scalar) parts:
\begin{equation}
{}^{(3)}\!R^{\alpha\beta} = {\frac {\overline{\mathcal M}}{2}}\,\eta^{\alpha\beta},\qquad 
{}^{(6)}\!R^{\alpha\beta} = {\frac {{\mathcal M}}{2}}\,\vartheta^\alpha\wedge\vartheta^\beta.\label{6Rab}
\end{equation}
\end{widetext}
By the direct inspection of (\ref{1R0a})-(\ref{6Rab}) we check the double duality properties of the irreducible curvature parts:
\begin{equation}\label{DDR}
{}^*({}^{(I)}\!R_{\alpha\beta}) = {\frac {K_I}{2}}\,\eta_{\alpha\beta\mu\nu}\,{}^{(I)}\!R^{\mu\nu},
\end{equation}
where $K_I = 1$ for $I = 1, 3, 5, 6$, and $K_I = -\,1$ for $I = 2, 4$. One can prove that (\ref{DDR}) is true not only for the spherically symmetric configurations but also in general case \cite{yno:2006}. These double duality properties will play a crucial role in the analysis of the validity of the generalized Birkhoff's theorem in the Poincar\'e gauge gravity.

\subsection{Weitzenb\"ock spacetime limit}

There is a special interest in the literature to the gravity models based on the teleparallelism when the Riemann-Cartan curvature is zero (Weitzenb\"ock spacetime). For the sake of completeness, here we describe the spherically symmetric field configuration for this case.

Assuming $\overline{f} = \overline{g} = \overline{k} = \overline{h} = 0$, we automatically have $\overline{\mu} = \overline{\nu} = 0$ and $\overline{U}{}^A = 0$, and the Riemann-Cartan curvature (\ref{RABS})-(\ref{RAbS}) becomes flat $R^{\alpha\beta} = 0$ when
\begin{equation}
\mu = 0,\qquad \nu = 0,\qquad U^A = 0.\label{flatS}
\end{equation}
Directly from the definitions (\ref{mumuS})-(\ref{U1S}), we find the general solution of the system of differential and algebraic equations (\ref{flatS}):
\begin{equation}
f = \dot{W},\quad g = W',\quad q = \cosh W,\quad p = -\,\sinh W,\label{fgkh0}
\end{equation}
where $W = W(t,r)$ is an arbitrary function \cite{MH}. 

\begin{widetext}

As a result, the local Lorentz connection (\ref{G01S})-(\ref{G03S}) reduces to
\begin{equation}\label{GflatS}
\Gamma_\alpha{}^\beta = \left(\begin{array}{c|c|c|c}
0 & dW & -\sinh Wd\theta & -\sinh W\sin\theta d\varphi \\ \hline
dW & 0 & \cosh Wd\theta & \cosh W\sin\theta d\varphi \\ \hline
-\sinh Wd\theta & - \cosh Wd\theta & 0 & -\cos\theta d\varphi \\ \hline 
-\sinh W\sin\theta d\varphi & -\cosh W\sin\theta d\varphi & \cos\theta d\varphi & 0
\end{array}\right),
\end{equation}
using the obvious matrix notation. It is straightforward to check that
\begin{equation}
\Gamma_\alpha{}^\beta = (\Lambda^{-1})^\beta{}_\gamma d \Lambda^\gamma{}_\alpha,\label{Gflat0}
\end{equation}
where the local Lorentz transformation $\Lambda^\alpha{}_\beta$ matrix is a product of two rotations and a boost $\Lambda = {\rm R_1R_2 L}$, with
\begin{equation}
{\rm R}_1 =  \left(\begin{array}{c|c|c|c}
1 & 0 & 0 & 0 \\ \hline 0 & 1 & 0 & 0 \\ \hline
0 & 0 & \cos\varphi & -\sin\varphi \\ \hline
0 & 0 & \sin\varphi & \cos\varphi \end{array}\right),\qquad
{\rm R_2} =  \left(\begin{array}{c|c|c|c}
1 & 0  & 0 & 0 \\ \hline
0 & \cos\theta & \sin\theta & 0 \\ \hline 0 & -\sin\theta & \cos\theta & 0 \\ \hline
0 & 0 & 0 & 1 \end{array}\right),\qquad
{\rm L} =  \left(\begin{array}{c|c|c|c}
\cosh W & \sinh W & 0 & 0 \\ \hline \sinh W & \cosh W & 0 & 0 \\ \hline
0 & 0 & 1 & 0 \\ \hline 0 & 0 & 0 & 1 \end{array}\right).
\end{equation}
\end{widetext}
Another case is described by $\overline{f} = \overline{g} = p = q = 0$, and
\begin{equation}
f = \dot{W},\quad g = W',\quad \overline{p} = \cosh W,\quad \overline{q} = \sinh W,\label{fgkh1}
\end{equation}
where as before $W = W(t,r)$ is an arbitrary function.

\section{Generalized Birkhoff theorem}\label{GBT}

After these preparations, we are in a position to analyse the generalized Birkhoff theorem in the Poincar\'e gauge gravity theory.

We begin by recalling that there are two versions of this theorem: the strong and the weak ones. It is worthwhile to explain this terminology which was proposed earlier in \cite{RauchGRG:1982,Rauch:1981,RauchPRD:1982} and may sound confusing, at least at the first sight. It seems reasonable to view a result derived under smaller number of assumptions as stronger than the same result obtained under more restrictive assumptions. In the context of our study of the Birkhoff theorem, the assumption of the spherical symmetry can be described either by $SO(3)$ rotation group, or by a larger (hence more restrictive) full rotation group $O(3)$. 

Accordingly, in the strong version of Birkhoff's theorem, the spherical symmetry is understood as the invariance under the pure rotations from the proper $SO(3)$ group. The corresponding symmetric connection configuration is described by the set (\ref{G01S})-(\ref{G03S}) of 8 arbitrary functions: $f, g, p, q$, and $\overline{f}, \overline{g}, \overline{p}, \overline{q}$. An equivalent description can be formulated in terms of the spherically symmetric torsion configuration that also encompasses eight variables (\ref{uv}) -- the components of the 1-forms $T, \overline{T}, V, \overline{V}$, see (\ref{u0S})-(\ref{tvS}).

In the weak version of Birkhoff's theorem, the spherical symmetry is understood as invariance under the full rotation group $O(3)$ which includes spatial reflections along with the pure rotations. Accordingly, in the weak version only parity-even objects are allowed, whereas the parity-odd variables are forbidden: $\overline{f} = \overline{g} = \overline{p} = \overline{q} = 0$, which reduces to 4 the number of connection components. This then yields $\overline{t}_A =\overline{v}_A = 0$ and hence $\overline{T} = \overline{V} = 0$, reducing the number of nontrivial torsion components also to 4.

To prove the generalized Birkhoff theorem, one needs to plug the spherically symmetric ansatz (\ref{cofSS}) and (\ref{G01S})-(\ref{G03S}) into the field equations (\ref{ERT1})-(\ref{ERT2}) and establish the conditions under which these field equations admit only solutions with the vanishing torsion and the Schwarzschild metric. Some of these conditions may restrict the coupling constants (hence, the structure of the Lagrangian), other conditions may impose constraints on the geometric properties of spacetime. Among the latter are: (i) an asymptotic flatness condition which requires that the metric (\ref{Sds}) approaches the Minkowski line element, i.e. $A \longrightarrow 1$, $B \longrightarrow 1$, $C \longrightarrow r$ in the limit of $r\longrightarrow\infty$, or (ii) an assumption of the vanishing scalar curvature $X = R = e_\alpha\rfloor e_\beta\rfloor R^{\alpha\beta} = 0$. 

In the earlier literature \cite{Rama,RauchGRG:1982,Rauch:1981,RauchPRD:1982,Neville:1978,Neville:1980,Cruz,OPZ}, only the parity-even class of models was analyzed with $\overline{a}_I = 0$, $\overline{b}_J = 0$. Here we consistently include the parity-odd sector into the consideration.

\subsection{Strong $SO(3)$ version}

In order to make the discussion more clear, we will explicitly mention the corresponding symmetry group in the name of a theorem, and thus instead of potentially confusing names for the ``strong'' and ``weak'' Birkhoff's theorem, we will speak of the ``strong $SO(3)$ version'' and the ``weak $O(3)$ version'', respectively. 

The strong $SO(3)$ version of the generalized Birkhoff theorem reads: the torsion-less Schwarzschild (Kottler, in general) spacetime is a unique solution of the vacuum Poincar\'e gauge field equations (\ref{ERT1})-(\ref{ERT2}) which is spherically symmetric in the sense of invariance under the proper rotation $SO(3)$ group.

It is important to stress that vanishing of the torsion is not an additional assumption on top of the spherical symmetry, but one finds that the torsion vanishes from the field equations.

\subsubsection{Strong $SO(3)$ version: case (SB1)}

Let us consider the class of models (\ref{LRT}) without the curvature-square terms,
\begin{equation}
b_I = 0,\qquad \overline{b}_I = 0,\qquad I = 1,\dots,6.\label{bb0}
\end{equation}
From (\ref{hR}) and (\ref{qaR}) we find $h_{\alpha\beta} = 0$ and $q^{(R)}_\alpha = 0$, and the second PG field equation (\ref{ERT2}) then reduces to an algebraic system for the torsion components, which can be recast into
\begin{eqnarray}
-\,a_0{}^*\left({}^{(1)}T^\alpha + 2\,{}^{(2)}T^\alpha + {\frac 12}{}^{(3)}T^\alpha\right) \nonumber\\
-\,\overline{a}{}_0\Big({}^{(1)}T^\alpha + {}^{(2)}T^\alpha + {}^{(3)}T^\alpha\Big) 
- h^\alpha = 0.\label{ECHQT}
\end{eqnarray}
Substituting (\ref{hT}), we derive
\begin{eqnarray}
-\,(a_0 + a_1){}^*({}^{(1)}T^\alpha) + (2a_0 - a_2){}^*({}^{(2)}T^\alpha)\nonumber\\
+\,\Big({\frac {a_0}2} - a_3\Big){}^*({}^{(3)}T^\alpha) - \Big(\overline{a}{}_0 +\overline{a}_1\Big)
{}^{(1)}T^\alpha\nonumber\\ - \Big(\overline{a}{}_0 + \overline{a}_2\Big)
{}^{(2)}T^\alpha - \Big(\overline{a}{}_0 + \overline{a}_3\Big){}^{(3)}T^\alpha = 0.\label{ECQT}
\end{eqnarray}
A direct inspection of this system, combined with its Hodge dual, shows that the vanishing torsion
$T^\alpha = 0$ is its only solution, provided
\begin{equation}\label{aanot}
\left. \begin{split} (a_0 + a_1)(2a_0 - a_2)(a_0 - 2a_3) \neq 0,\\
(\overline{a}{}_0 +\overline{a}_1)(\overline{a}{}_0 + \overline{a}_2)
(\overline{a}{}_0 + \overline{a}_3) \neq 0.\end{split}\right\}
\end{equation}
Then the first PG field equation  (\ref{ERT1}) reduces to the usual Einstein equation
\begin{equation}
{\frac {a_0}2}\eta_{\alpha\beta\gamma}\wedge \widetilde{R}^{\beta\gamma} - \lambda_0\eta_\alpha = 0.\label{E0}
\end{equation}
Hereafter we denote the Riemannian objects by the tilde. In particular, the Levi-Civita connection $\widetilde{\Gamma}{}^{\alpha\beta}$ is torsion-free, $\widetilde{D}\vartheta^\alpha = d\vartheta^\alpha + \widetilde{\Gamma}_\beta{}^\alpha\wedge\vartheta^\beta = 0$, and $\widetilde{R}{}^{\alpha\beta} = d\widetilde{\Gamma}{}^{\alpha\beta} + \widetilde{\Gamma}_\gamma{}^\beta\wedge\widetilde{\Gamma}{}^{\alpha\gamma}$ is the corresponding Riemannian curvature 2-form. 

Since the standard Birkhoff theorem holds for (\ref{E0}), we thus have demonstrated the validity of the generalized Birkhoff theorem for the family of PG models (\ref{bb0}), (\ref{aanot}).

\subsubsection{Strong $SO(3)$ version: case (SB2)}

Let us consider a family of PG models where quadratic curvature terms are included in the Lagrangian (\ref{LRT}) with the coupling constants chosen as
\begin{equation}\label{bb4}
b_1 = b_5 = -\,b_2 = -\,b_4,\qquad \overline{b}_1 = \overline{b}_2 = \overline{b}_4 = \overline{b}_5.
\end{equation}
Then, by making use of the double duality properties (\ref{DDR}), for (\ref{hR}) we have
\begin{eqnarray}
h_{\alpha\beta} &=& {\frac {b_1}{2}}\,\eta_{\alpha\beta\mu\nu}\,R^{\mu\nu} + \overline{b}_1\,R_{\alpha\beta}
- b_{1-3}{}^{*(3)}\!R_{\alpha\beta} \nonumber\\
&& -\,b_{1-6}{}^{*(6)}\!R_{\alpha\beta} - \overline{b}_{1-3}\,(
{}^{(3)}\!R_{\alpha\beta} + {}^{(6)}\!R_{\alpha\beta}).\label{hR1}
\end{eqnarray}
Here we introduced a convenient compact notation for the linear combinations of the coupling
constants:
\begin{equation}\label{bpm}
b_{I\pm J} := b_I \pm b_J,\qquad \overline{b}_{I\pm J} := \overline{b}_I \pm \overline{b}_J\,.
\end{equation}

Making an {\it additional assumption} that the curvature scalars ${\mathcal M}$ and $\overline{\mathcal M}$ are constant, we then find the covariant derivative
\begin{eqnarray}
Dh_{\alpha\beta} &=& -\,{\frac {1}{2}}\left(b_{1-6}{\mathcal M} + \overline{b}_{1-3}\overline{\mathcal M}\right)
D\eta_{\alpha\beta} \nonumber\\
&& -\,{\frac 12}\left(\overline{b}_{1-3}{\mathcal M} - b_{1-3}\overline{\mathcal M}\right)
D(\vartheta_\alpha\wedge\vartheta_\beta),\label{DhR1}
\end{eqnarray}
where we used (\ref{6Rab}), and the covariant derivative of the first two terms in (\ref{hR1}) vanish in view of the Bianchi identity $DR_{\alpha\beta} = 0$. Substituting (\ref{DhR1}) into the second PG field equation (\ref{ERT2}), the latter yields 
\begin{eqnarray}
-\,(a^{\rm eff}_0 + a_1){}^*({}^{(1)}T^\alpha) + (2a^{\rm eff}_0 - a_2){}^*({}^{(2)}T^\alpha)\nonumber\\
+\,\Big({\frac {a^{\rm eff}_0}2} - a_3\Big){}^*({}^{(3)}T^\alpha) - \Big(\overline{a}{}^{\rm eff}_0 +\overline{a}_1\Big)
{}^{(1)}T^\alpha\nonumber\\ - \Big(\overline{a}{}^{\rm eff}_0 + \overline{a}_2\Big)
{}^{(2)}T^\alpha - \Big(\overline{a}{}^{\rm eff}_0 + \overline{a}_3\Big){}^{(3)}T^\alpha = 0,\label{ECQT2}
\end{eqnarray}
where we introduced the effective coupling constants
\begin{eqnarray}\label{a0e2}
a^{\rm eff}_0 &=& a_0 + \ell_\rho^2(b_{1-6}{\mathcal M} + \overline{b}_{1-3}\overline{\mathcal M}),\\
\overline{a}{}^{\rm eff}_0 &=& \overline{a}{}_0 + \ell_\rho^2(\overline{b}_{1-3}{\mathcal M}
- b_{1-3}\overline{\mathcal M}).\label{oa0e2}
\end{eqnarray}
The vanishing torsion $T^\alpha = 0$ is the only solution of (\ref{ECQT2}), provided
\begin{equation}\label{aanot2}
\left. \begin{split} (a^{\rm eff}_0 + a_1)(2a^{\rm eff}_0 - a_2)(a^{\rm eff}_0 - 2a_3) \neq 0,\\
(\overline{a}{}^{\rm eff}_0 +\overline{a}_1)(\overline{a}{}^{\rm eff}_0 + \overline{a}_2)
(\overline{a}{}^{\rm eff}_0 + \overline{a}_3) \neq 0,\end{split}\right\}
\end{equation}
and then the first PG field equation (\ref{ERT1}) again reduces to the usual Einstein equation (\ref{E0}) and the validity of generalized Birkhoff theorem is therefore established.

\subsubsection{Strong $SO(3)$ version: case (SB3)}

For a sub-family of PG models (\ref{bb4}) when the choice (\ref{bb4}) of coupling constants is further specified to 
\begin{equation}\label{bb44}
\left.\begin{split}
b_1 = b_3 = b_5 = b_6 = -\,b_2 = -\,b_4 ,\\
\overline{b}_1 = \overline{b}_2 = \overline{b}_3 = \overline{b}_4 = \overline{b}_5 = \overline{b}_6,
\end{split}\right\}
\end{equation}
the 2-form (\ref{hR1}) is simplified to 
\begin{eqnarray}\label{hR4}
h_{\alpha\beta} = {\frac {b_1}{2}}\,\eta_{\alpha\beta\mu\nu}\,R^{\mu\nu} + \overline{b}_1\,R_{\alpha\beta},
\end{eqnarray}
so that $D h_{\alpha\beta} = 0$, and the second field equation (\ref{ERT2}) yields (\ref{ECQT}). The latter, as we already know, admits the vanishing torsion as a unique solution, provided the coupling constants satisfy (\ref{aanot}), and the generalized Birkhoff theorem is established because $q^{(R)}_\alpha = 0$ for (\ref{bb44}) and therefore the first PG field equation (\ref{ERT1}) again reduces to GR's (\ref{E0}).

\subsubsection{Strong $SO(3)$ version: case (SB4)}

Let us conclude the analysis of the strong Birkhoff theorem by considering the class of models (\ref{LRT}) without the torsion-square terms,
\begin{equation}
a_I = 0,\qquad \overline{a}_I = 0,\qquad I = 1,2,3,\label{aa0}
\end{equation}
when the curvature-square coupling constants satisfy (\ref{bb4}), and in addition, are restricted by the relations
\begin{eqnarray}
\overline{a}_0\overline{b}_{1-3} + a_0 b_{1-3} = 0,\quad
\overline{a}_0b_{1-6} - a_0\overline{b}_{1-3} = 0.\label{abab} 
\end{eqnarray}
In view of (\ref{aa0}), we have $h_\alpha = 0$, and computing the trace of the first PG field equation (\ref{ERT1}), $\vartheta^\alpha\wedge{\mathcal E}_\alpha = 0$, we get
\begin{equation}
a_0{\mathcal M} + \overline{a}_0\overline{\mathcal M} - {\frac {2\lambda_0}{3}} = 0.\label{aaMM}
\end{equation}
On the other hand, $h_{\alpha\beta}$, which is still given by (\ref{hR1}), with the help of (\ref{abab}) reduces to    
\begin{eqnarray}
h_{\alpha\beta} = {\frac {b_1}{2}}\,\eta_{\alpha\beta\mu\nu}\,R^{\mu\nu} + \overline{b}_1\,R_{\alpha\beta}\nonumber\\
-\,{\frac {a_0{\mathcal M} + \overline{a}_0\overline{\mathcal M}}{a_0}}
\left(b_{1-6}\eta_{\alpha\beta} + \overline{b}_{1-3}\,\vartheta_\alpha\wedge\vartheta_\beta\right).\label{hR3}
\end{eqnarray}
With an account of (\ref{aaMM}), the second PG field equation (\ref{ERT2}) then becomes a special case of (\ref{ECQT2}) in which we should insert (\ref{aa0}). The effective coupling constants remain the same (\ref{a0e2}) and (\ref{oa0e2}). As a result, we discover the vanishing torsion $T^\alpha = 0$ as the only solution of (\ref{ECQT2}), provided
\begin{equation}
a^{\rm eff}_0 \neq 0,\qquad \overline{a}{}^{\rm eff}_0 \neq 0,\label{aanot3}
\end{equation}
and the generalized Birkhoff theorem is therefore valid, since the first PG field equation (\ref{ERT1}) again reduces to the usual Einstein equation (\ref{E0}).

\subsection{Weak $O(3)$ version}

The weak $O(3)$ version of the generalized Birkhoff theorem reads: the torsion-less Schwarzschild (Kottler, in general) spacetime is a unique solution of the vacuum Poincar\'e gauge field equations (\ref{ERT1})-(\ref{ERT2}), which is spherically symmetric in the sense of invariance under the full rotation $O(3)$ group, when spatial reflections are included in addition to the proper rotations. 

As before, an important remark is that vanishing of the torsion is not an additional assumption on top of the spherical symmetry, but one finds that the torsion vanishes from the field equations. 

In order to study the weak $O(3)$ version of the generalized Birkhoff theorem, we put $\overline{f} = \overline{g} = \overline{p} = \overline{q} = 0$ which are forbidden by the spatial reflection symmetry. Therefore, $\overline{t}_A =\overline{v}_A = 0$ and hence $\overline{T} = \overline{V} = 0$, and also $\overline{U}{}^A = 0$. As a result, we find $\overline{\mathcal Q} = \overline{\mathcal P} = \overline{\mathcal L} = \overline{\mathcal N} = \overline{\mathcal K} = \overline{\mathcal M} = 0$. Consequently, from (\ref{ST23}), (\ref{2R0a})-(\ref{2Rab}), (\ref{6Rab}) we conclude that one torsion irreducible part and two of the irreducible curvature parts are trivial
\begin{equation}\label{TRR0}
{}^{(3)}\!T^\alpha = 0,\quad {}^{(2)}\!R_{\alpha\beta} = 0,\quad {}^{(3)}\!R_{\alpha\beta} = 0.
\end{equation}

Similarly to the strong version of the generalized Birkhoff theorem, the weak version can be established for several families of quadratic PG models.

\subsubsection{Weak $O(3)$ version: case (WB1)}

Let us consider the class of models (\ref{LRT}) without the curvature-square terms, 
\begin{equation}
b_I = 0,\qquad \overline{b}_I = 0,\qquad I = 1,\dots,6.\label{Wbb0}
\end{equation}
From we have $h_{\alpha\beta} = 0$ and $q^{(R)}_\alpha = 0$, and the second PG field equation (\ref{ERT2}) then reduces to 
\begin{eqnarray}
-\,(a_0 + a_1){}^*({}^{(1)}T^\alpha) + (2a_0 - a_2){}^*({}^{(2)}T^\alpha)\nonumber\\
-\,\Big(\overline{a}{}_0 +\overline{a}_1\Big){}^{(1)}T^\alpha - \Big(\overline{a}{}_0 + \overline{a}_2\Big)
{}^{(2)}T^\alpha = 0.\label{WECQT}
\end{eqnarray}
The vanishing torsion $T^\alpha = 0$ is its only solution, provided
\begin{equation}\label{Waanot}
(a_0 + a_1)(2a_0 - a_2) \neq 0,\quad
(\overline{a}{}_0 +\overline{a}_1)(\overline{a}{}_0 + \overline{a}_2) \neq 0.
\end{equation}
The first PG field equation (\ref{ERT1}) then reduces to the usual Einstein equation (\ref{E0}), which demonstrates the validity of the generalized Birkhoff theorem for the family of PG models (\ref{bb0}), (\ref{aanot}).

\subsubsection{Weak $O(3)$ version: case (WB2)}

Let us consider a family of PG models where quadratic curvature terms are included in the Lagrangian (\ref{LRT}) with the coupling constants chosen as
\begin{equation}\label{Wbb4}
b_1 = b_5 = -\,b_4,\qquad \overline{b}_1 = \overline{b}_4 = \overline{b}_5.
\end{equation}
Then, by making use of the double duality properties (\ref{DDR}), for (\ref{hR}) we have
\begin{eqnarray}
h_{\alpha\beta} &=& {\frac {b_1}{2}}\,\eta_{\alpha\beta\mu\nu}\,R^{\mu\nu} + \overline{b}_1\,R_{\alpha\beta}\nonumber\\
&& -\,b_{1-6}{}^{*(6)}\!R_{\alpha\beta} - \overline{b}_{1-3}\,{}^{(6)}\!R_{\alpha\beta}.\label{WhR1}
\end{eqnarray}
Under an {\it additional assumption} that the curvature scalar ${\mathcal M}$ is constant, we then find the covariant derivative
\begin{eqnarray}
Dh_{\alpha\beta} = -\,{\frac {{\mathcal M}}{2}}\left[b_{1-6}\,D\eta_{\alpha\beta} +
\overline{b}_{1-3}\,D(\vartheta_\alpha\wedge\vartheta_\beta)\right],\label{WDhR1}
\end{eqnarray}
and hence the second PG field equation (\ref{ERT2}) reduces to
\begin{eqnarray}
-\,(a^{\rm eff}_0 + a_1){}^*({}^{(1)}T^\alpha) + (2a^{\rm eff}_0 - a_2){}^*({}^{(2)}T^\alpha)\nonumber\\
-\, \Big(\overline{a}{}^{\rm eff}_0 +\overline{a}_1\Big){}^{(1)}T^\alpha - \Big(\overline{a}{}^{\rm eff}_0
+ \overline{a}_2\Big){}^{(2)}T^\alpha = 0,\label{WECQT2}
\end{eqnarray}
with the effective coupling constants
\begin{eqnarray}
a^{\rm eff}_0 &=& a_0 + b_{1-6}\,\ell_\rho^2{\mathcal M},\label{Wa0e2}\\
\overline{a}{}^{\rm eff}_0 &=& \overline{a}{}_0 + \overline{b}_{1-3}\,\ell_\rho^2{\mathcal M}.\label{Woa0e2}
\end{eqnarray}
The vanishing torsion $T^\alpha = 0$ is the only solution of (\ref{WECQT2}), provided
\begin{equation}\label{Waanot2}
(a^{\rm eff}_0 + a_1)(2a^{\rm eff}_0 - a_2) \neq 0,\quad
(\overline{a}{}^{\rm eff}_0 +\overline{a}_1)(\overline{a}{}^{\rm eff}_0 + \overline{a}_2) \neq 0,
\end{equation}
and then the first PG field equation (\ref{ERT1}) again reduces to the usual Einstein equation (\ref{E0}) and the validity of generalized Birkhoff theorem is therefore established.

\subsubsection{Weak $O(3)$ version: case (WB3)}

For a sub-family of PG models (\ref{Wbb4}) when the choice (\ref{Wbb4}) of coupling constants is further specified to 
\begin{equation}\label{Wbb44}
b_1 = b_5 = b_6 = -\,b_4 ,\qquad
\overline{b}_1 = \overline{b}_4 = \overline{b}_5 = \overline{b}_6,
\end{equation}
the 2-form (\ref{WhR1}) is simplified to 
\begin{eqnarray}\label{WhR4}
h_{\alpha\beta} = {\frac {b_1}{2}}\,\eta_{\alpha\beta\mu\nu}\,R^{\mu\nu} + \overline{b}_1\,R_{\alpha\beta},
\end{eqnarray}
so that $D h_{\alpha\beta} = 0$, and the second field equation (\ref{ERT2}) yields (\ref{WECQT}). The latter, as we already know, admits the vanishing torsion as a unique solution, provided the coupling constants satisfy (\ref{Waanot}), and the generalized Birkhoff theorem is established because $q^{(R)}_\alpha = 0$ for (\ref{Wbb44}) and therefore the first PG field equation (\ref{ERT1}) again reduces to GR's (\ref{E0}).

\subsubsection{Weak $O(3)$ version: case (WB4)}

Similarly to the analysis of the strong Birkhoff theorem, let us consider the class of models (\ref{LRT}) without the torsion-square terms,
\begin{equation}
a_1 = a_2 = 0,\qquad \overline{a}_1 = \overline{a}_2 = 0,\label{Waa0}
\end{equation}
whereas the curvature-square coupling constants satisfy (\ref{Wbb4}). In view of (\ref{Waa0}), we have $h_\alpha = 0$, and computing the trace of the first PG field equation (\ref{ERT1}), $\vartheta^\alpha\wedge{\mathcal E}_\alpha = 2(3a_0{\mathcal M} - 2\lambda_0)\eta = 0$, we get
\begin{equation}
{\mathcal M} = {\frac {2\lambda_0}{3a_0}}.\label{Mconst}
\end{equation}
Since $h_{\alpha\beta}$ is still given by (\ref{WhR1}), with an account of (\ref{Mconst}) its covariant derivative is the same (\ref{WDhR1}), and the second PG field equation (\ref{ERT2}) becomes a special case of (\ref{WECQT2}) in which we should insert (\ref{Waa0}). The effective coupling constants remain the same (\ref{Wa0e2}) and (\ref{Woa0e2}). Consequently, the vanishing torsion $T^\alpha = 0$ is the only solution of (\ref{WECQT2}), provided
\begin{equation}
a^{\rm eff}_0 \neq 0,\qquad \overline{a}{}^{\rm eff}_0 \neq 0,\label{Waanot3}
\end{equation}
and the generalized Birkhoff theorem is therefore valid, since the first PG field equation (\ref{ERT1}) again reduces to the usual Einstein equation (\ref{E0}).

\subsubsection{Weak $O(3)$ version: cases (WB5), (WB6), (WB7)}

Let us study in greater detail the case (\ref{Waa0}) when $a_1 = a_2 = 0$ and $\overline{a}_1 = \overline{a}_2 = 0$, however without apriori fixing the values of the curvature-square coupling constants $b_I$, $\overline{b}_I$. Then $h_\alpha = 0$ and $q_\alpha^{(T)} = 0$, and the PG field equations (\ref{ERT1}) and (\ref{ERT2}) read in vacuum 
\begin{eqnarray}\label{LHS1}
{\mathcal E}_\alpha = {\frac {a_0}2}\eta_{\alpha\beta\gamma}\wedge R^{\beta\gamma} + \overline{a}{}_0
R_{\alpha\beta}\wedge\vartheta^\beta \nonumber\\ -\,\lambda_0\eta_\alpha + \ell_\rho^2\,q^{(R)}_\alpha = 0,\\
{\mathcal C}_{\alpha\beta} = a_0\,\eta_{\alpha\beta\gamma}\wedge T^{\gamma} + \overline{a}{}_0\left(
T_\alpha\wedge\vartheta_\beta - T_\beta\wedge\vartheta_\alpha \right) \nonumber\\
- \,2\ell_\rho^2\,Dh_{\alpha\beta} = 0.\label{LHS2}
\end{eqnarray}
For the trace of the first field equation (\ref{LHS1}), we found the constant scalar curvature (\ref{Mconst}).

Before we consider the general case, let us analyse the special case when
\begin{equation}
{\mathcal Q} = {\mathcal P} = {\mathcal L} = {\mathcal N} = 0.\label{special}
\end{equation}
Then from (\ref{4R0a})-(\ref{5Rab}) we conclude that ${}^{(4)}\!R_{\alpha\beta} = {}^{(5)}\!R_{\alpha\beta} = 0$, and in view of (\ref{TRR0}) the curvature has a simple structure
\begin{equation}
R_{\alpha\beta} = {}^{(1)}\!R_{\alpha\beta} + {}^{(6)}\!R_{\alpha\beta}.\label{Rspec}
\end{equation}
As a result, the 2-form (\ref{hR}) is given by (\ref{WhR1}), its covariant derivative $Dh_{\alpha\beta}$ is given by (\ref{WDhR1}) with an account of (\ref{Mconst}), and we then find for the second PG field equation (\ref{LHS2}):
\begin{equation}
a^{\rm eff}_0\eta_{\alpha\beta\gamma}\wedge T^{\gamma} + \overline{a}{}^{\rm eff}_0
\left(T_\alpha\wedge\vartheta_\beta - T_\beta\wedge\vartheta_\alpha \right) = 0,\label{LHS2spec}
\end{equation}
where the effective coupling constants are (\ref{Wa0e2}) and (\ref{Woa0e2}). The algebraic equation (\ref{LHS2spec}) has a unique trivial torsion $T^\alpha = 0$ under the condition (\ref{Waanot3}). Indeed, contracting (\ref{LHS2spec}) with the Levi-Civita tensor $\eta^{\alpha\beta\mu\nu}$, one finds
\begin{equation}
a^{\rm eff}_0\,(T_\alpha\wedge\vartheta_\beta - T_\beta\wedge\vartheta_\alpha)
- \overline{a}{}^{\rm eff}_0\eta_{\alpha\beta\gamma}\wedge T^{\gamma} = 0,\label{LHS2spec2}
\end{equation}\\
and combining this with (\ref{LHS2spec}), we prove that the solution for the torsion is trivial under the condition (\ref{Waanot3}). 

Turning to the first field equation, one can straightforwardly verify that $q^{(R)}_\alpha = 0$ for the curvature (\ref{Rspec}), and the field equation (\ref{LHS1}) reduces to (\ref{E0}) the standard Einstein equation for the Riemannian case for which the generalized Birkhoff theorem is valid.

We will call (\ref{Waanot3}) a {\it primary condition} and from now on will assume that it is satisfied. For the class of models without the cosmological constant $\lambda_0 = 0$ (hence ${\mathcal M} = 0$), the primary condition reduces to the requirement $a_0 \neq 0, \overline{a}_0\neq 0$ which excludes the purely curvature quadratic models. We will therefore assume that $a_0 \neq 0$ and $\overline{a}_0\neq 0$ and taking into account (\ref{Mconst}), we write down the primary condition explicitly as
\begin{equation}\label{prim}
\left(a_0 + {\frac {2\ell_\rho^2\lambda_0}{3a_0}}\,b_{1 - 6}\right)
\left(\overline{a}{}_0 + {\frac {2\ell_\rho^2\lambda_0}{3a_0}}\,\overline{b}_{1-6}\right) \neq 0.
\end{equation}

\begin{widetext}
Now, let us consider the general case, without assuming (\ref{special}) in advance. Inserting the spherically symmetric ansatz for the coframe (\ref{cofSS}) and connection (\ref{G01S})-(\ref{G03S}) into the first field equation (\ref{LHS1}), we find explicitly:
\begin{eqnarray}
{\mathcal E}_{\hat 0} &=& \eta_{\hat 0}\left\{ {\frac 12}a_0(3{\mathcal M} - 2{\mathcal L}
- {\mathcal N}) - \lambda_0 + {\frac 12}\ell_\rho^2\bigl[2b_{4 + 5}{\mathcal Q}{\mathcal P}
+\,2b_{1 + 4}{\mathcal K}({\mathcal N} - {\mathcal L}) +
b_{4 + 6}{\mathcal M}(2{\mathcal L} + {\mathcal N})\bigr]\right\}\nonumber\\
&& +\,\eta_{\hat 1}\left\{ a_0({\mathcal P} - {\mathcal Q}) + \ell_\rho^2\bigl[ b_{1 + 4}{\mathcal K}
{\mathcal P} - b_{4 + 6}{\mathcal M}{\mathcal P} 
-\,b_{1 - 5}{\mathcal Q}{\mathcal K} - b_{4 + 5} {\mathcal Q}{\mathcal L}
- b_{5 - 6}{\mathcal Q}{\mathcal M}\bigr]\right\},\label{ES0}\\
{\mathcal E}_{\hat 1} &=& \eta_{\hat 1}\left\{ {\frac 12}a_0(3{\mathcal M} + 2{\mathcal L}
- {\mathcal N}) - \lambda_0 + {\frac 12}\ell_\rho^2\bigl[-\,2b_{4 + 5}{\mathcal Q}{\mathcal P}
+\,2b_{1 + 4}{\mathcal K}({\mathcal N} + {\mathcal L}) +
b_{4 + 6}{\mathcal M}(- 2{\mathcal L} + {\mathcal N})\bigr]\right\}\nonumber\\
&& +\,\eta_{\hat 0}\left\{ -\,a_0({\mathcal P} + {\mathcal Q}) + \ell_\rho^2\bigl[
-\,b_{1 + 4}{\mathcal K}{\mathcal P} + b_{4 + 6}{\mathcal M}{\mathcal P} 
-\,b_{1 - 5}{\mathcal Q}{\mathcal K} + b_{4 + 5} {\mathcal Q}{\mathcal L}
- b_{5 - 6}{\mathcal Q}{\mathcal M}\bigr]\right\}.\label{ES1}\\
{\mathcal E}_{\hat 2} &=& \eta_{\hat 2}\left\{ {\frac 12}a_0(3{\mathcal M} + {\mathcal N})
- \lambda_0 + {\frac 12}\ell_\rho^2\bigl[-\,2b_{1 + 4}{\mathcal K}{\mathcal N}
- b_{4 + 6}{\mathcal M}{\mathcal N}\bigr]\right\}
-\,\eta_{\hat 3}\,{\mathcal Q}\left\{\overline{a}_0 + \ell_\rho^2\left[\overline{b}_{1 - 5}
 {\mathcal K} + \overline{b}_{5 - 6}{\mathcal M}\right]\right\},\label{ES2}\\
{\mathcal E}_{\hat 3} &=& \eta_{\hat 3}\left\{ {\frac 12}a_0(3{\mathcal M} + {\mathcal N})
- \lambda_0 + {\frac 12}\ell_\rho^2\bigl[-\,2b_{1 + 4} {\mathcal K}{\mathcal N}
- b_{4 + 6}{\mathcal M}{\mathcal N}\bigr]\right\}
+\,\eta_{\hat 2}\,{\mathcal Q}\left\{\overline{a}_0 + \ell_\rho^2\left[\overline{b}_{1 - 5}
{\mathcal K} + \overline{b}_{5 - 6}{\mathcal M}\right]\right\}.\label{ES3}
\end{eqnarray}
\end{widetext}
The trace (\ref{Mconst}) can be verified directly from (\ref{ES0})-(\ref{ES3}). With an account of
(\ref{Mconst}), the vacuum field equations (\ref{ES0})-(\ref{ES3}) are recast into the system 
\begin{eqnarray}
{\mathcal Q}\left\{\overline{a}_0 + \ell_\rho^2\left[\overline{b}_{1 - 5}
{\mathcal K} + \overline{b}_{5 - 6}{\mathcal M}\right]\right\} = 0,\label{EF5}\\
a_0{\mathcal N} - \ell_\rho^2\left[ 2b_{1 + 4}{\mathcal K}{\mathcal N}
+ b_{4 + 6}{\mathcal M}{\mathcal N}\right] = 0,\label{EF1a}\\
a_0{\mathcal L} - \ell_\rho^2\left[ b_{4 + 5}{\mathcal Q}{\mathcal P}
- b_{1 + 4}{\mathcal K}{\mathcal L} + b_{4 + 6}{\mathcal M}{\mathcal L}
\right] = 0,\label{EF2a}\\
a_0{\mathcal P} - \ell_\rho^2\left[ b_{4 + 5}{\mathcal Q}{\mathcal L}
- b_{1 + 4}{\mathcal K}{\mathcal P} + b_{4 + 6}{\mathcal M}{\mathcal P}
\right] = 0,\label{EF3a}\\
{\mathcal Q}\left\{a_0 + \ell_\rho^2\left[b_{1 - 5}
{\mathcal K} + b_{5 - 6}{\mathcal M}\right]\right\} = 0.\label{EF4a}
\end{eqnarray}

From the equations (\ref{EF4a}) and (\ref{EF5}) we conclude that ${\mathcal Q} = 0$, except for an exotic case when the coupling constants satisfy
\begin{eqnarray}
a_0 + \ell_\rho^2\left[b_{1 - 5}{\mathcal K} + b_{5 - 6}{\mathcal M}\right] &=& 0,\\
\overline{a}_0 + \ell_\rho^2\left[\overline{b}_{1 - 5}
{\mathcal K} + \overline{b}_{5 - 6}{\mathcal M}\right] &=& 0, 
\end{eqnarray}
which is possible only when 
\begin{equation}
a_0\,\overline{b}_{1 - 5} - \overline{a}_0\,b_{1 - 5} +
{\frac {2\ell_\rho^2\lambda_0}{3a_0}}\left[\,\overline{b}_{1 - 5}\,b_{5 - 6} 
- b_{1 - 5}\,\overline{b}_{5 - 6}\,\right] = 0.\label{exc}  
\end{equation}

The analysis of the second field equation (\ref{LHS2}) is simplified greatly by making use of the double-duality properties of the curvature. Namely, the crucial observation is that we can recast the 2-form $h_{\alpha\beta}$, defined in (\ref{hR}), into
\begin{equation}
h_{\alpha\beta} = -\,{\frac {b_4}{2}}\,\eta_{\alpha\beta\mu\nu}\,R^{\mu\nu} + \overline{b}_4\,R_{\alpha\beta}
+ h^{\rm eff}_{\alpha\beta},\label{hR2}
\end{equation}
where we introduced the effective quantity
\begin{eqnarray}
h^{\rm eff}_{\alpha\beta} &=& {}^*\!\left(b_{1+4}\,{}^{(1)}\!R_{\alpha\beta} + b_{5+4}\,{}^{(5)}\!R_{\alpha\beta}
+ b_{6+4}\,{}^{(6)}\!R_{\alpha\beta}\right)\nonumber\\
&+& \,\overline{b}_{1-4}\,{}^{(1)}\!R_{\alpha\beta} + \overline{b}_{5-4}\,{}^{(5)}\!R_{\alpha\beta}
+ \overline{b}_{6-4}\,{}^{(6)}\!R_{\alpha\beta}.\label{hReff}
\end{eqnarray}
The derivation is based on a straightforward use of the properties (\ref{DDR}). The advantage of the representation (\ref{hR2}) is that we identically have
\begin{equation}\label{DhDh}
Dh_{\alpha\beta} \equiv Dh^{\rm eff}_{\alpha\beta},
\end{equation}
in view of the Bianchi identity $DR_{\alpha\beta} = 0$. 

\begin{widetext}
The second field equation (\ref{LHS2}) read in components
\begin{align}
{\mathcal C}_{\hat{0}\hat{2}} &= {\mathcal H}_0\eta_{\hat{2}} - \overline{\mathcal H}_0\eta_{\hat{3}} = 0, &
{\mathcal C}_{\hat{0}\hat{3}} &= {\mathcal H}_0\eta_{\hat{3}} + \overline{\mathcal H}_0\eta_{\hat{2}} = 0, &
{\mathcal C}_{\hat{0}\hat{1}} &= {\mathcal F}_0\eta_{\hat{1}} - {\mathcal F}_1\eta_{\hat{0}} = 0,\label{SF01}\\
{\mathcal C}_{\hat{1}\hat{2}} &= {\mathcal H}_1\eta_{\hat{2}} - \overline{\mathcal H}_1\eta_{\hat{3}} = 0, & 
{\mathcal C}_{\hat{1}\hat{3}} &= {\mathcal H}_1\eta_{\hat{3}} + \overline{\mathcal H}_1\eta_{\hat{2}} = 0, &
{\mathcal C}_{\hat{2}\hat{3}} &= \overline{\mathcal F}_0\eta_{\hat{1}} - \overline{\mathcal F}_1\eta_{\hat{0}} = 0,\label{SF23}
\end{align}
and making use of (\ref{hReff}), (\ref{DhDh}) and (\ref{1R0a})-(\ref{6Rab}), we find explicitly 
\begin{eqnarray}
{\mathcal F}_0 &=& \ell_\rho^2\Bigl\{ {\frac {2b_{1+4}\dot{\mathcal K} + b_{6+4}\dot{\mathcal M}}{A}}
+ {\frac {2(3b_{1+4}{\mathcal K}p + b_{5+4}{\mathcal Q}q)}{C}}\Bigr\}
+ 2\Bigl(v_{\hat 0} - {\frac 13}t_{\hat 0}\Bigr)\left[ -\,a_0 + \ell_\rho^2
(2b_{1+4}{\mathcal K} + b_{6+4}{\mathcal M})\right],\label{CF0}\\
{\mathcal F}_1 &=& \ell_\rho^2\Bigl\{ {\frac {2b_{1+4}{\mathcal K}' + b_{6+4}{\mathcal M}'}{B}}
- {\frac {2(3b_{1+4}{\mathcal K}q + b_{5+4}{\mathcal Q}p)}{C}}\Bigr\}
+ 2\Bigl(v_{\hat 1} - {\frac 13}t_{\hat 1}\Bigr)\left[ -\,a_0 + \ell_\rho^2
(2b_{1+4}{\mathcal K} + b_{6+4}{\mathcal M})\right],\label{CF1}\\
\overline{\mathcal F}_0 &=& \ell_\rho^2\Bigl\{ {\frac {2\overline{b}_{1-4}\dot{\mathcal K}
+ \overline{b}_{6-4}\dot{\mathcal M}}{A}} + {\frac {2(3\overline{b}_{1-4}{\mathcal K}p
+ \overline{b}_{5-4}{\mathcal Q}q)}{C}}\Bigr\}
+ 2\Bigl(v_{\hat 0} - {\frac 13}t_{\hat 0}\Bigr)\left[ -\,\overline{a}_0 + \ell_\rho^2
(2\overline{b}_{1-4}{\mathcal K} + \overline{b}_{6-4}{\mathcal M})\right],\label{CTF0}\\
\overline{\mathcal F}_1 &=& \ell_\rho^2\Bigl\{ {\frac {2\overline{b}_{1-4}{\mathcal K}'
+ \overline{b}_{6-4}{\mathcal M}'}{B}} - {\frac {2(3\overline{b}_{1-4}{\mathcal K}q
+ \overline{b}_{5-4}{\mathcal Q}p)}{C}}\Bigr\}
+ 2\Bigl(v_{\hat 1} - {\frac 13}t_{\hat 1}\Bigr)\left[ -\,\overline{a}_0 + \ell_\rho^2
(2\overline{b}_{1-4}{\mathcal K} + \overline{b}_{6-4}{\mathcal M})\right],\label{CTF1} \\
{\mathcal H}_0 &=& \ell_\rho^2\Bigl\{{\frac {-\,b_{1+4}\dot{\mathcal K} + b_{6+4}\dot{\mathcal M}}{A}}
- {\frac {b_{5+4}{\mathcal Q}'}{B}} - {\frac {3b_{1+4}{\mathcal K}p - b_{5+4}{\mathcal Q}q}C}\Bigr\}
+ t_{\hat 1}\ell_\rho^2b_{5+4}{\mathcal Q} + t_{\hat 0}[ a_0
+ \ell_\rho^2\left(b_{1+4}{\mathcal K} - b_{6+4}{\mathcal M}\right)],\label{CH0}\\
{\mathcal H}_1 &=& \ell_\rho^2\Bigl\{{\frac {-\,b_{1+4}{\mathcal K}' + b_{6+4}{\mathcal M}'}{B}}
- {\frac {b_{5+4}\dot{\mathcal Q}}{A}}  + {\frac {3b_{1+4}{\mathcal K}q - b_{5+4}{\mathcal Q}p}C}\Bigr\}
+ t_{\hat 0}\ell_\rho^2b_{5+4}{\mathcal Q} + t_{\hat 1}[ a_0
+ \ell_\rho^2(b_{1+4}{\mathcal K} - b_{6+4}{\mathcal M})],\label{CH1} \\
\overline{\mathcal H}_0 &=& \ell_\rho^2\Bigl\{{\frac {\overline{b}_{1-4}{\mathcal K}' 
- \overline{b}_{6-4}{\mathcal M}'}{B}} + {\frac {\overline{b}_{5-4}\dot{\mathcal Q}}{A}}
- {\frac {3\overline{b}_{1-4}{\mathcal K}q + \overline{b}_{5-4}{\mathcal Q}p}C}\Bigr\}
- t_{\hat 0}\ell_\rho^2\overline{b}_{5-4}{\mathcal Q} - t_{\hat 1}[ \overline{a}_0
+ \ell_\rho^2\left(\overline{b}_{1-4}{\mathcal K} - \overline{b}_{6-4}{\mathcal M}\right)],\label{CTH0}\\
\overline{\mathcal H}_1 &=& \ell_\rho^2\Bigl\{{\frac {\overline{b}_{1-4}\dot{\mathcal K}
- \overline{b}_{6-4}\dot{\mathcal M}}{A}} + {\frac {\overline{b}_{5-4}{\mathcal Q}'}{B}}
+ {\frac {3\overline{b}_{1-4}{\mathcal K}p - \overline{b}_{5-4}{\mathcal Q}q}C}\Bigr\}
- t_{\hat 1}\ell_\rho^2\overline{b}_{5-4}{\mathcal Q} - t_{\hat 0}[\overline{a}_0
+ \ell_\rho^2(\overline{b}_{1-4}{\mathcal K} - \overline{b}_{6-4}{\mathcal M})].\label{CTH1} 
\end{eqnarray}
\end{widetext}

After these preliminaries, we are in a position to demonstrate that Birkhoff's theorem is valid for the following three cases (WB5), (WB6) and (WB7):

{\sl Case (WB5)}. Assuming the primary condition (\ref{Waanot3}), (\ref{prim}), the family of PG models specified by
\begin{equation}
b_1 = -\,b_4 = b_5.\label{B1}
\end{equation}

If, in addition to the primary condition, we assume the {\it secondary} conditions
\begin{eqnarray}
a_0 - {\frac {2\ell_\rho^2\lambda_0}{3a_0}}\,(b_4 + b_6) &\neq& 0,\label{sec1}\\
a_0 - {\frac {2\ell_\rho^2\lambda_0}{3a_0}}\,(2b_1 + 3b_4 + b_6) &\neq& 0,\label{sec2}
\end{eqnarray}
we find two more possibilities:

{\sl Case (WB6)}. Lagrangians with the coupling constants 
\begin{equation}
b_1 = b_5,\qquad b_4 \neq -\,b_5.\label{B2}
\end{equation}

{\sl Case (WB7)}:
\begin{equation}
b_4 = -\,b_5,\qquad b_1 \neq b_5.\label{B3}
\end{equation}

The proof is straightforward. {\sl Case (WB5)}: When the coupling constant satisfy (\ref{B1}), or equivalently, $b_{1+4} = b_{4+5} = 0$ and $b_{1-5} = 0$, the set of equations (\ref{EF1a})-(\ref{EF4a}) reduces to
\begin{eqnarray}
{\mathcal N}\left(a_0 + \ell_\rho^2{\mathcal M}b_{1 - 6}\right) &=& 0,\label{EF1b}\\
{\mathcal L}\left(a_0 + \ell_\rho^2{\mathcal M}b_{1 - 6}\right) &=& 0,\label{EF2b}\\
{\mathcal P}\left(a_0 + \ell_\rho^2{\mathcal M}b_{1 - 6}\right) &=& 0,\label{EF3b}\\
{\mathcal Q}\left(a_0 + \ell_\rho^2{\mathcal M}b_{1 - 6}\right) &=& 0,\label{EF4b}
\end{eqnarray}
and in view of the primary condition (\ref{Waanot3}), we find ${\mathcal Q} = {\mathcal P} = {\mathcal N} = {\mathcal L} = 0$. Thus the special case (\ref{special}) is recovered for which the validity of the Birkhoff theorem was already established.

{\sl Case (WB6)}. Assuming (\ref{B2}), eq. (\ref{EF4a}) is recast into
\begin{equation}
{\mathcal Q}\left\{a_0 + \ell_\rho^2{\mathcal M}b_{1 - 6}\right\} = 0,\label{EF4c}
\end{equation}
and hence ${\mathcal Q} = 0$ due to the primary condition (\ref{Waanot3}). As a result, the system (\ref{EF1a})-(\ref{EF3a}) reduces to 
\begin{eqnarray}
{\mathcal N}\left\{-\,a_0 + \ell_\rho^2\left[2b_{1 + 4}{\mathcal K}
+ b_{4 + 6}{\mathcal M}\right]\right\} &=& 0,\label{EF1c}\\
{\mathcal L}\left\{-\,a_0 + \ell_\rho^2\left[ - b_{1 + 4}{\mathcal K}
+ b_{4 + 6}{\mathcal M}\right]\right\} &=& 0,\label{EF2c}\\
{\mathcal P}\left\{-\,a_0 + \ell_\rho^2\left[ - b_{1 + 4}{\mathcal K}
+ b_{4 + 6}{\mathcal M}\right]\right\} &=& 0.\label{EF3c}
\end{eqnarray}
Therefore, either ${\mathcal P} = {\mathcal N} = {\mathcal L} = 0$, and we again recover the special case (\ref{special}), or we face the two subcases:
\begin{equation}\label{B2a}
-\,a_0 + \ell_\rho^2\left[2b_{1 + 4}{\mathcal K} + b_{4 + 6}{\mathcal M}\right] = 0,
\end{equation}
or
\begin{equation}\label{B2b}
-\,a_0 + \ell_\rho^2\left[ - b_{1 + 4}{\mathcal K} + b_{4 + 6}{\mathcal M}\right] = 0.
\end{equation}

{\sl Subcase WB6.1}. If we have (\ref{B2a}), then under the secondary condition (\ref{sec1}), 
\begin{equation}\label{B2a1}
\ell_\rho^2b_{1 + 4}{\mathcal K} = {\frac 12}\left(a_0 - \ell_\rho^2{\mathcal M}b_{4 + 6}\right) \neq 0.
\end{equation}
But in this subcase the field equations (\ref{CF0}) and (\ref{CF1}), that is ${\mathcal F}_0 = 0$ and
${\mathcal F}_1 = 0$, then yield
\begin{equation}\label{pq0}
p = 0,\qquad q = 0,
\end{equation}
which means that $U^A = 0$, see (\ref{U0S}) and (\ref{U1S}). By making use of (\ref{LP}), we therefore conclude that ${\mathcal K} = {\mathcal M}$ and substituting this into (\ref{B2a}), we find
\begin{equation}\label{B2a2}
-\,a_0 + \ell_\rho^2{\mathcal M}\left(2b_{1 + 4} + b_{4 + 6}\right) = 0,
\end{equation}
which contradicts the secondary condition (\ref{sec2}).

{\sl Subcase WB6.2}. Assuming (\ref{B2b}) under the secondary condition (\ref{sec1}) we find
\begin{equation}\label{B2b1}
\ell_\rho^2b_{1 + 4}{\mathcal K} = -\,\left(a_0 - \ell_\rho^2{\mathcal M}b_{4 + 6}\right) \neq 0,
\end{equation}
and using the field equations (\ref{CH0}) and (\ref{CH1}), i.e., ${\mathcal H}_0 = 0$ and ${\mathcal H}_1 = 0$, we again end up with (\ref{pq0}). This again yields $U^A = 0$ via (\ref{U0S}) and (\ref{U1S}), and by using (\ref{LP}), we have ${\mathcal K} = {\mathcal M}$ as in the previous subcase. Substituting this into (\ref{B2b}), we find
\begin{equation}\label{B2b2}
-\,a_0 + \ell_\rho^2{\mathcal M}\left( -\,b_{1 + 4} + b_{4 + 6}\right) = 
-\left(a_0 + \ell_\rho^2{\mathcal M}b_{1 - 6}\right) = 0,  
\end{equation}
which thus contradicts the secondary condition (\ref{sec1}).

{\sl Case (WB7)} is similar. Under the assumption (\ref{B3}), the system (\ref{EF1a})-(\ref{EF4a}) reduces to
\begin{eqnarray}
{\mathcal N}\left\{-\,a_0 + \ell_\rho^2\left[2b_{1 + 4}{\mathcal K}
+ b_{4 + 6}{\mathcal M}\right]\right\} &=& 0,\label{EF1d}\\
{\mathcal L}\left\{-\,a_0 + \ell_\rho^2\left[ - b_{1 + 4}{\mathcal K}
+ b_{4 + 6}{\mathcal M}\right]\right\} &=& 0,\label{EF2d}\\
{\mathcal P}\left\{-\,a_0 + \ell_\rho^2\left[ - b_{1 + 4}{\mathcal K}
+ b_{4 + 6}{\mathcal M}\right]\right\} &=& 0,\label{EF3d}\\
{\mathcal Q}\left\{-\,a_0 + \ell_\rho^2\left[ - b_{1 + 4}{\mathcal K}
+ b_{4 + 6}{\mathcal M}\right]\right\} &=& 0.\label{EF4d}
\end{eqnarray}
Therefore, either ${\mathcal Q} = {\mathcal P} = {\mathcal N} = {\mathcal L} = 0$ and we recover the special case (\ref{special}) for which the validity of the Birkhoff theorem was already established, or we have to consider the two subcases (\ref{B2a}) and (\ref{B2b}). The corresponding analysis is verbatim the same as in the subcases WB6.1 and WB6.2 above.

\section{On the physical meaning of primary and secondary conditions}

In order to reach a better understanding of the primary and secondary conditions (\ref{prim}), (\ref{sec1}), (\ref{sec2}), let us go beyond the spherical symmetry. A Riemannian spacetime which arises as a solution of the vacuum Einstein equation (\ref{E0}) with a cosmological constant is called an Einstein space. A thorough investigation of Einstein spaces can be found in the book of Petrov \cite{Petrov}. If we put the torsion equal zero in the general quadratic PG field equations (\ref{ERT1})-(\ref{ERT2}), one can straightforwardly verify that Einstein spaces (\ref{E0}) are exact solutions of the latter. 

However, a much stronger and more nontrivial result concerns the uniqueness of Einstein spaces as torsionless solutions of PG field equations. Namely, one can prove \cite{OPZ,yno:2006,yno:2018,Fairchild,Debney} that Einstein spaces (\ref{E0}) are the only torsionless solutions of the general quadratic PG field equations (\ref{ERT1})-(\ref{ERT2}) for an arbitrary PG Lagrangian (\ref{LRT}), excluding the {\it three exceptional cases} when the coupling constants satisfy one of the following equalities
\begin{eqnarray}
a_0 + {\frac {2\ell_\rho^2\lambda_0}{3a_0}}\,(b_1 - b_6) &=& 0,\label{ex1}\\
a_0 - {\frac {2\ell_\rho^2\lambda_0}{3a_0}}\,(b_4 + b_6) &=& 0,\label{ex2}\\
a_0 - {\frac {2\ell_\rho^2\lambda_0}{3a_0}}\,(2b_1 + 3b_4 + b_6) &=& 0.\label{ex3}
\end{eqnarray}
The field equations of the PG models, belonging to one of these exceptional cases, admit torsionless solutions which are {\it not} Einstein spaces. As a specific example we can mention the Stephenson-Kilmister-Yang (SKY) gravity theory \cite{Stephenson,Kilmister,Yang} which is known to have non-Einsteinian vacuum solutions \cite{Ni,Pavelle,Thompson1,Thompson2}. 

In other words, the primary and secondary conditions (\ref{prim}), (\ref{sec1}), (\ref{sec2}) are not only necessary to ensure the validity of the generalized Birkhoff theorem for the spherically symmetric solutions of the quadratic PG models, they are also sufficient conditions to prohibit (\ref{ex1})-(\ref{ex3}) and thereby to guarantee that the only torsionless solutions of the Poincar\'e field equations are Einstein spaces.

\section{Vacuum solutions violating Birkhoff's theorem}\label{violate}

All quadratic PG models admit Schwarzschild (or in general, Kottler, when the cosmological constant is nontrivial) spherically symmetric solution without torsion. However, other spherically symmetric solutions, besides the torsionless Schwarzschild/Kottler, may exist in those PG models which do not belong to one of the families described in Sec.~\ref{GBT}. Here we consider such a case and explicitly construct a spherically symmetric solution with nontrivial torsion. Let us specialize to the Lagrangian (\ref{LRT}) where the torsion-square part is described by the coupling constants
\begin{equation}\label{aCV}
\left.\begin{split}
a_1 = -\,a_0,\quad a_2 = 2a_0,\quad a_3 = {\frac 12}a_0,\\
\overline{a}_1 = -\,\overline{a}_0,\quad \overline{a}_2 = -\,\overline{a}_0,\quad \overline{a}_3 = -\,\overline{a}_0,
\end{split}\right\}
\end{equation}
which directly violates (\ref{aanot}). At the same time, we fix the structure of the curvature-square part of the Lagrangian (\ref{LRT}) by the following choice of the coupling constants:
\begin{equation}\label{bCV}
\left.\begin{split}
b_3 = - \,b_2,\quad b_5 = - \,{\frac {b_2}{3}},\qquad b_1 = b_4 = b_6 = 0,\\
\overline{b}_3 = \overline{b}_2,\quad \overline{b}_5 = {\frac {\overline{b}_2}{3}},\qquad 
\overline{b}_1 = \overline{b}_4 = \overline{b}_6 = 0.
\end{split}\right\}
\end{equation}
Therefore, this model is characterized by the two coupling parameters $b_2$ and $\overline{b}_2$, in addition to $a_0$ and $\overline{a}_0$.

There are two geometrical identities \cite{yno:2006,yno:2018} which hold for the choice of the constants (\ref{aCV}):
\begin{eqnarray}
{\frac {a_0}2}\eta_{\alpha\beta\gamma}\wedge R^{\beta\gamma} + \overline{a}_0R_{\alpha\beta}\wedge\vartheta^\beta\nonumber\\
+\,q^{(T)}_\alpha - Dh_\alpha \equiv {\frac {a_0}2}\eta_{\alpha\beta\gamma}\wedge \widetilde{R}{}^{\beta\gamma},\label{id1}
\end{eqnarray}
\begin{equation}
a_0\,\eta_{\alpha\beta\gamma}\wedge T^{\gamma} + 2\overline{a}_0\,T_{[\alpha}\wedge\vartheta_{\beta]}
+ 2h_{[\alpha}\wedge\vartheta_{\beta]} \equiv 0.\label{id2}
\end{equation}
Both these relations are valid for all configurations of Poincar\'e gauge fields (not only spherically symmetric) irrespectively of their dynamics. 

As a result, the vacuum PG field equations (\ref{ERT1}) and (\ref{ERT2}) reduce to 
\begin{eqnarray}
{\frac {a_0}2}\eta_{\alpha\beta\gamma}\wedge \widetilde{R}{}^{\beta\gamma} - \lambda_0\eta_\alpha
+ \ell_\rho^2\,q^{(R)}_\alpha &=& 0,\label{ECV1}\\
- 2\ell_\rho^2\,Dh_{\alpha\beta} &=& 0.\label{ECV2}
\end{eqnarray}
Recall that the tilde denotes the Riemannian objects. In view of (\ref{bCV}), we have explicitly
\begin{eqnarray}
h_{\alpha\beta} &=& b_2\,{}^*\!\Bigl({}^{(2)}\!R_{\alpha\beta} - {}^{(3)}\!R_{\alpha\beta} - {\frac 13}
{}^{(5)}\!R_{\alpha\beta}\Bigr)\nonumber\\
&& + \,\overline{b}_2\Bigl({}^{(2)}\!R_{\alpha\beta} + {}^{(3)}\!R_{\alpha\beta} + {\frac 13}
{}^{(5)}\!R_{\alpha\beta}\Bigr).\label{habCV}
\end{eqnarray}

The subsequent derivations are based on the methods developed in \cite{GW1,GW2,yno:2019}. Let us further specify the general spherically symmetric ansatz (\ref{cofSS}), (\ref{G01S})-(\ref{G03S}) and consider the static (no time dependence) configuration 
\begin{eqnarray}
B = {\frac 1A},\quad C = r,\label{ABCV}\\
f = 0,\quad g = {\frac {A'}{A}},\quad p = -\,q = {\frac A2},\label{fgVC}\\ \label{pqCV}
{\frac {\overline{f}}{A}} + {\frac {\overline{g}}{B}} = 0,\quad \overline{p} = \overline{q} = 0.
\end{eqnarray}
The prime $^\prime$ denotes derivative with respect to the radial coordinate $r$.
This simplifies the torsion 2-form $T^\alpha$ to  
\begin{eqnarray}
T^{\hat 0} = T^{\hat 1} = - \,A'\,\vartheta^{\hat 0}\wedge\vartheta^{\hat 1},\label{T01CV}\\
T^{\hat 2} = -\,{\frac {A}{2r}}\,k\wedge\vartheta^{\hat 2}
+ {\frac {\overline{f}}{A}}\,k\wedge\vartheta^{\hat 3},\label{T2CV} \\
T^{\hat 3} = -\,{\frac {A}{2r}}\,k\wedge\vartheta^{\hat 3}
- {\frac {\overline{f}}{A}}\,k\wedge\vartheta^{\hat 2},\label{T3CV}
\end{eqnarray}
where we introduced a null 1-form
\begin{equation}
k = \vartheta^{\hat 0} - \vartheta^{\hat 1}.\label{kCV} 
\end{equation}
Accordingly, the torsion 1-forms (\ref{VVTT}) reduce to
\begin{eqnarray}
T = \Bigl(A' + {\frac Ar}\Bigr)k,\qquad \overline{T} = -2\,{\frac {\overline{f}}{A}}\,k,\label{TCV}\\
V = {\frac 13}\Bigl(A' - {\frac Ar}\Bigr)k,\qquad \overline{V} = {\frac {\overline{f}}{3A}}\,k.\label{VCV}
\end{eqnarray}
As a result, the torsion invariant (\ref{TTS}) vanishes in view of the null property $k\wedge {}^*k = 0$ of the 1-form (\ref{kCV}).

At the same time, the spherically symmetric curvature 2-form (\ref{RABS})-(\ref{RAbS}) also considerably simplifies because the coefficients (\ref{mumuS})-(\ref{Ut1S}) reduce to
\begin{eqnarray}
\mu = 0,\qquad \overline{\mu} = 0,\label{mumuCV}\\
\nu = {\frac {1}{r^2}},\qquad \overline{\nu} = -\,\overline{f}{}',\label{nunuCV}\\
U^A = 0,\qquad \overline{U}{}^{\hat 0} = -\,\overline{U}{}^{\hat 1} =
-\,{\frac {\overline{f}}{2r}}\,k.\label{UUCV}
\end{eqnarray}
Consequently, we have
\begin{equation}\label{UTCV}
e_A\rfloor \overline{U}{}^A = -\,{\frac {\overline{f}}{r}},\qquad \vartheta_A\wedge
\overline{U}{}^A = {\frac {\overline{f}}{r}}\,\vartheta^{\hat 0}\wedge\vartheta^{\hat 1}.
\end{equation}
Inserting (\ref{mumuCV})-(\ref{UTCV}) into (\ref{MN})-(\ref{OLP}), we derive explicitly
\begin{eqnarray}
&&{\mathcal M} = {\frac {1}{3r^2}},\quad {\mathcal K} = {\frac {1}{3r^2}},\quad {\mathcal N}
= -\,{\frac {1}{r^2}},\label{MKNCV}\\
&&{\mathcal Q} = 0,\qquad {\mathcal L} = 0,\qquad {\mathcal P} = 0,\label{QLPCV}\\
&&\overline{\mathcal M} = -\,{\frac {\overline{f}{}^\prime}{3}} - {\frac {2\overline{f}}{3r}},\quad
\overline{\mathcal K} = -\,{\frac {\overline{f}{}^\prime}{3}} + {\frac {\overline{f}}{3r}},\quad
\overline{\mathcal N} = \overline{f},\label{OMKNCV}\\
&&\overline{\mathcal Q} = {\frac {\overline{f}}{r}},\qquad \overline{\mathcal L} = 0,
\qquad \overline{\mathcal P} = 0.\label{OQLPCV}
\end{eqnarray}
As we see, the torsion and the curvature are determined by the two unknown functions $A = A(r)$ and $\overline{f} = \overline{f}(r)$. It remains to check whether one can solve the Poincar\'e field equations (\ref{ECV1}) and (\ref{ECV2}) for some $A, \overline{f}$.

The spherically symmetric configuration of the torsion (\ref{T01CV})-(\ref{T3CV}) and the Riemann-Cartan curvature (\ref{MKNCV})-(\ref{OQLPCV}) described above has a remarkable property that the covariant derivative of the 2-form (\ref{habCV}) vanishes
\begin{equation}
Dh_{\alpha\beta} = 0,\label{DhCV}
\end{equation}
provided the function $\overline{f}$ has the form of a ``Coulomb potential'' 
\begin{equation}
\overline{f} = {\frac {\tau_0}{r}},\label{fCV}
\end{equation}
with an arbitrary constant $\tau_0$. Accordingly, we obtain an exact vacuum solution of the second field equation (\ref{ECV2}).

Another remarkable property of $h_{\alpha\beta}$ is that the 3-form (\ref{qaR}) reads
\begin{equation}\label{qaCV}
q^{(R)}_\alpha = -\,{\frac {Q^2}{r^4}}\left[\delta^{\hat 0}_\alpha\,\eta_{\hat 0} + \delta^{\hat 1}_\alpha
\,\eta_{\hat 1} - \delta^{\hat 2}_\alpha\,\eta_{\hat 2} - \delta^{\hat 3}_\alpha\,\eta_{\hat 3}\right],
\end{equation}
where the constant
\begin{equation}
Q^2 = {\frac {2b_2\tau_0^2 + \overline{b}_2\tau_0}3}.\label{QCV}
\end{equation}
Substituting this into (\ref{ECV1}), we immediately verify that the first field equation is solved for
\begin{equation}\label{RNCV}
A^2 = 1 - {\frac {2m}{r}} + {\frac {\ell_\rho^2Q^2}{a_0r^2}} - {\frac {\lambda_0}{3a_0}}r^2.
\end{equation}

We thus recover a Reissner-Nordstr\"om-de Sitter solution where the integration constant $m$ can be interpreted as the mass of a point-like source. However, unlike in the genuine Reissner-Nordstr\"om gravitational field, here the effective ``electric charge'' of the source is generated by the geometry of spacetime. Moreover, although we formally introduced $Q^2$ as a quadratic quantity, it is obvious that (\ref{QCV}) is not necessarily positive, with its value and sign depending on the constants $b_2, \overline{b}_2$ and $\tau_0$.

This spherically symmetric vacuum solution of the PG theory (which extends the Cembranos-Valcarcel results \cite{CV1,CV2} to a much wider family of Lagrangians) represents an explicit example of the violation of Birkhoff's theorem.

\section{Discussion and conclusions}

In this paper, we have revisited the Birkhoff theorem for the Poincar\'e gauge gravity theory. In the class of quadratic PG models with the most general Lagrangian (\ref{LRT}) which includes both the parity-even and parity-odd terms, we have established the validity of the generalized Birkhoff theorem for the families of models (SB1)-(SB4) and (WB1)-(WB7) in the strong $SO(3)$ and the weak $O(3)$ versions, respectively. In the strong version, the spherical symmetry is understood as the invariance under the pure rotations from the proper $SO(3)$ group, while in the weak version of Birkhoff's theorem, the spherical symmetry is understood as invariance under the full rotation group $O(3)$ which includes spatial reflections along with the pure rotations. 

With an account of a nontrivial cosmological constant, the Birkhoff theorem in both versions states that the only locally spherically symmetric solution of the PG field equations in vacuum (\ref{ERT1})-(\ref{ERT2}) is either the Schwarzschild-(anti)de Sitter (Kottler) metric or the Nariai (Bertotti-Kasner) spacetime without torsion. The vanishing of the torsion is not an additional assumption on top of the spherical symmetry, but this is a consequence of the field equations.

It is important to notice that the generalized Birkhoff theorem is not valid for the most general Lagrangian (\ref{LRT}), which was known already for the parity-even class of models studied earlier in the literature \cite{Rama,RauchGRG:1982,Rauch:1981,RauchPRD:1982,Neville:1978,Neville:1980,Cruz,OPZ}. In order to demonstrate the violation of the generalized Birkhoff theorem, in Sec.~\ref{violate} we explicitly construct a spherically symmetric solution with torsion for the PG model which does not belong to the families (SB1)-(SB4) and (WB1)-(WB7).

As a comment to the primary and the secondary conditions (\ref{prim}) and (\ref{sec1}), (\ref{sec2}), we should note that they impose only mild restrictions on the structure of the PG Lagrangian. From the technical point of view, these conditions merely rule out some special values of the cosmological constant, and when $\lambda_0 = 0$ these conditions reduce just to the requirement that the Lagrangian should necessarily contain linear curvature terms with nonvanishing $a_0$ and $\overline{a}_0$. This explains why the Birkhoff theorem is violated in the purely quadratic models such as the SKY gravity.

An interesting question arises whether our methods and results can be extended to other gravitational theories. The metric-affine gravity (MAG) \cite{MAG,Blag:2013} is the nearest generalization of the Poincar\'e gauge gravity, in which the spacetime geometry includes, in addition to the torsion, the nonmetricity as another post-Riemannian geometrical structure. Based on the recent analysis of the spherical symmetry in the metric-affine geometry \cite{MH1,MH2}, one can study the validity of the Birkhoff theorem for MAG models with the Yang-Mills type quadratic Lagrangians along the lines of the PG approach discussed above. On the other hand, it may be of interest going beyond the quadratic Lagrangians. In the recent times, the teleparallel gravity and especially the modified $f(T)$ and $f(T,B)$ models \cite{Capozziello:2020,Cai:2016} attract much of attention in relation with the discussion of the dark matter and the dark energy problems in the modern astrophysics and cosmology. Since such models are dynamically related via conformal transformations to scalar-tensor theories \cite{Capo,Cai:2016}, it seems reasonable to study a possibility of combining our approach with the conformal/scalar-tensor methods. We plan to address these open issues in a future research. 

The results obtained contribute to the understanding of the dynamical structure of the Poincar\'e gravity theory. The analysis of the validity of the generalized Birkhoff theorem helps to reduce the large number of the coupling constants in the general quadratic Lagrangian (\ref{LRT}) to a set that determines a class of physically viable models which are consistent with Einstein's GR at large distances. Other criteria such as the unitarity and stability (absence of ghost and tachyon modes in the particle spectrum) impose additional restrictions on the coupling constants \cite{Hay,Sez1,Sez2,Nev1,Nev2,Ark,Kuh,Bai,VCV,Percacci:2020,Jimenez,Kara,BC,MH}, which should be combined with our findings.

\begin{acknowledgments}
I am grateful to Friedrich W.\ Hehl (Cologne) for the constant support and encouragement, deep questions and helpful comments. This work was partially supported by the Russian Foundation for Basic Research (Grant No. 18-02-40056-mega).
\end{acknowledgments}

\appendix
\section{Irreducible decompositions}\label{appA}

The torsion 2-form can be decomposed into the three irreducible pieces, $T^{\alpha}={}^{(1)}T^{\alpha} + {}^{(2)}T^{\alpha} + {}^{(3)}T^{\alpha}$:
\begin{eqnarray}
{}^{(1)}T^{\alpha}&=& T^{\alpha}-{}^{(2)}T^{\alpha} - {}^{(3)}T^{\alpha},\label{iT1}\\
{}^{(2)}T^{\alpha} &=& {\frac 13}\vartheta^{\alpha}\wedge T,\label{iT2}\\
{}^{(3)}T^{\alpha}&=& -\,{\frac 13}{}^*(\vartheta^{\alpha}\wedge\overline{T}),\label{iT3}
\end{eqnarray}
where the 1-forms of the torsion {\it trace} and {\it axial trace} are defined as
\begin{equation}\label{TT}
T = e_\nu\rfloor T^\nu,\qquad \overline{T} = {}^*(T^{\nu}\wedge\vartheta_{\nu}).
\end{equation}

The Riemann-Cartan curvature 2-form is decomposed $R^{\alpha\beta} = 
\sum_{I=1}^6\,{}^{(I)}\!R^{\alpha\beta}$ into the 6 irreducible parts 
\begin{eqnarray}
{}^{(2)}\!R^{\alpha\beta} &=& -\,{}^*(\vartheta^{[\alpha}\wedge
\overline{\Psi}{}^{\beta]}),\label{curv2}\\
{}^{(3)}\!R^{\alpha\beta} &=& -\,{\frac 1{12}}\,{}^*(\overline{X}
\,\vartheta^\alpha\wedge\vartheta^\beta),\label{curv3}\\
{}^{(4)}\!R^{\alpha\beta} &=& -\,\vartheta^{[\alpha}\wedge\Psi^{\beta]},\label{curv4}
\end{eqnarray}
\begin{eqnarray}
{}^{(5)}\!R^{\alpha\beta} &=& -\,{\frac 12}\vartheta^{[\alpha}\wedge e^{\beta]}
\rfloor(\vartheta^\gamma\wedge X_\gamma),\label{curv5}\\
{}^{(6)}\!R^{\alpha\beta} &=& -\,{\frac 1{12}}\,X\,\vartheta^\alpha\wedge
\vartheta^\beta,\label{curv6}\\
{}^{(1)}\!R^{\alpha\beta} &=& R^{\alpha\beta} -  
\sum\limits_{I=2}^6\,{}^{(I)}R^{\alpha\beta},\label{curv1}
\end{eqnarray}
where we denoted
\begin{eqnarray}
X^\alpha := e_\beta\rfloor R^{\alpha\beta},\qquad X := e_\alpha\rfloor X^\alpha,\label{WX1}\\
\overline{X}{}^\alpha := {}^*(R^{\beta\alpha}\wedge\vartheta_\beta),\qquad
\overline{X} := e_\alpha\rfloor \overline{X}{}^\alpha,\label{WX2}\\
\Psi_\alpha := X_\alpha - {\frac 14}\,\vartheta_\alpha\,X - {\frac 12}
\,e_\alpha\rfloor (\vartheta^\beta\wedge X_\beta),\label{Psia}\\
\overline{\Psi}{}_\alpha := \overline{X}{}_\alpha - {\frac 14}\,\vartheta_\alpha
\,\overline{X} - {\frac 12}\,e_\alpha\rfloor (\vartheta^\beta\wedge 
\overline{X}{}_\beta).\label{Phia}
\end{eqnarray}

\end{document}